\documentclass[prd,reprint,twocolumn,superscriptaddress,nofootinbib,preprintnumbers]{revtex4-1}
\usepackage{amsmath,amssymb,latexsym,times}
\usepackage[utf8]{inputenc}

\usepackage{slashed}
\usepackage{float}
\usepackage{graphicx}
\usepackage{grffile}
\usepackage[normalem]{ulem}
\usepackage{dcolumn}
\usepackage[dvipsnames]{xcolor}
\usepackage{booktabs}
\usepackage{multirow}
\usepackage{amssymb,amsmath}
\usepackage{todonotes}
\usepackage{enumitem}
\usepackage{hyperref}
 \hypersetup{
     colorlinks=true,
     linkcolor=blue,
     citecolor = blue,
     urlcolor=blue,
     }
 \usepackage{xspace}
 \usepackage{booktabs}

\xspaceaddexceptions{]\}}
\usepackage{todonotes}

\usepackage[T1]{fontenc}
\usepackage{ae,aecompl}

\newcommand{\nside}{\ifmmode N_{\mathrm{side}}\else $N_{\mathrm{side}}$\fi}
\newcommand{\npix}{\ifmmode n_{\mathrm{pix}}\else $n_{\mathrm{pix}}$\fi}
\newcommand{\Npix}{\ifmmode N_{\mathrm{pix}}\else $n_{\mathrm{pix}}$\fi}
\newcommand{\lmin}{\ifmmode \ell_{\mathrm{min}}\else $\ell_{\mathrm{min}}$\fi}
\newcommand{\lmax}{\ifmmode \ell_{\mathrm{max}}\else $\ell_{\mathrm{max}}$\fi}

\renewcommand{\vec}[1]{{\mathbf{#1}}}

\newcommand{\sect}[1]{Sect.~\ref{#1}\xspace}

\newcommand{\app}[1]{Appendix~\ref{#1}\xspace}

\newcommand{\fig}[1]{Fig.~\ref{#1}\xspace}
\newcommand{\figs}[1]{Figs.~\ref{#1}\xspace}
\newcommand{\eq}[1]{Eq.~(\ref{#1})\xspace}
\newcommand{\eqs}[1]{Eqs.~(\ref{#1})\xspace}
\newcommand{\tab}[1]{Table~\ref{#1}\xspace}

\DeclareMathOperator{\erfc}{erfc}

\newcommand{\nn}{\nonumber\\}

\usepackage{lineno}

\newcommand{\lneqb}{\begin{linenomath*}}
\newcommand{\lneqe}{\end{linenomath*}}

\newcommand{\lcdm}{\ensuremath{\Lambda\text{CDM}}\xspace}
\newcommand{\wcdm}{\ensuremath{w\text{CDM}}\xspace}
\newcommand{\mpp}{\ensuremath{3\times 2\text{pt}}\xspace}
\newcommand{\cosmosis}{{\sc CosmoSIS}\xspace}
\newcommand{\halofit}{{\sc halofit}\xspace}

\newcommand{\yone}{Y1-3\ensuremath{\times}2pt\xspace}
\newcommand{\metacal}{{\sc Metacalibration}\xspace}
\newcommand{\redmagic}{{\sc redMaGiC}\xspace}

\newcommand{\plin}{\ensuremath{P_{\rm lin}}\xspace}
\newcommand{\plingrow}{\ensuremath{P_{\rm lin}^{\rm grow}}\xspace}
\newcommand{\plingeo}{\ensuremath{P_{\rm lin}^{\rm geo}}\xspace}
\newcommand{\params}{\ensuremath{\Theta}\xspace}
\newcommand{\paramsgrow}{\ensuremath{\Theta^{\rm grow}}\xspace}
\newcommand{\paramsgeo}{\ensuremath{\Theta^{\rm geo}}\xspace}
\newcommand{\om}{\ensuremath{\Omega_m}\xspace}
\newcommand{\omgrow}{\ensuremath{\Omega_m^{\rm grow}}\xspace}
\newcommand{\omgeo}{\ensuremath{\Omega_m^{\rm geo}}\xspace}
\newcommand{\wgrow}{\ensuremath{w^{\rm grow}}\xspace}
\newcommand{\wgeo}{\ensuremath{w^{\rm geo}}\xspace}
\newcommand{\mnu}{\ensuremath{\sum m_{\nu}}\xspace}

\newcommand{\sigeight}{\ensuremath{\sigma_8}\xspace}
\newcommand{\seight}{\ensuremath{S_8}\xspace}
\newcommand{\fsig}{\ensuremath{f\sigma_8}\xspace}

\newcommand{\chisq}{\ensuremath{\chi^2}\xspace}

\newcommand{\as}{\ensuremath{A_{\rm s}}\xspace}
\newcommand{\aia}{\ensuremath{A_{\rm IA}}\xspace}

\begin{document}

\title{DES Y1 results: Splitting growth and geometry to test \lcdm}

\author{J.~Muir}
\email[Corresponding author: ]{jlmuir@stanford.edu}
\affiliation{Kavli Institute for Particle Astrophysics \& Cosmology, P. O. Box 2450, Stanford University, Stanford, CA 94305, USA}
\author{E.~Baxter}
\affiliation{Department of Physics and Astronomy, Watanabe 416, 2505 Correa Road, Honolulu, HI 96822}
\author{V.~Miranda}
\affiliation{Department of Astronomy/Steward Observatory, University of Arizona, 933 North Cherry Avenue, Tucson, AZ 85721-0065, USA}
\author{C.~Doux}
\affiliation{Department of Physics and Astronomy, University of Pennsylvania, Philadelphia, PA 19104, USA}
\author{A.~Fert\'e}
\affiliation{Jet Propulsion Laboratory, California Institute of Technology, 4800 Oak Grove Dr., Pasadena, CA 91109, USA}
\author{C.~D.~Leonard}
\affiliation{School of Mathematics, Statistics and Physics, Newcastle University, Newcastle upon Tyne, NE1 7RU, UK}
\author{D.~Huterer}
\affiliation{Department of Physics, University of Michigan, Ann Arbor, MI 48109, USA}
\author{B.~Jain}
\affiliation{Department of Physics and Astronomy, University of Pennsylvania, Philadelphia, PA 19104, USA}
\author{P.~Lemos}
\affiliation{Department of Physics \& Astronomy, University College London, Gower Street, London, WC1E 6BT, UK}
\author{M.~Raveri}
\affiliation{Kavli Institute for Cosmological Physics, University of Chicago, Chicago, IL 60637, USA}
% ...

\collaboration{DES Collaboration}

\date{\today}

\label{firstpage}

\begin{abstract}
  We analyze Dark Energy Survey (DES) data to constrain a cosmological model where a subset of parameters --- focusing on $\Omega_m$ --- are split into versions associated with structure growth (e.g. $\Omega_m^{\rm grow}$) and expansion history (e.g. $\Omega_m^{\rm geo}$). Once the parameters have been specified for the $\Lambda$CDM cosmological model, which includes general relativity as a theory of gravity, it uniquely predicts the evolution of both geometry (distances) and the growth of structure over cosmic time. Any inconsistency between measurements of geometry and growth could therefore indicate a breakdown of that model. Our growth-geometry split approach therefore serves as both a (largely) model-independent test for beyond-$\Lambda$CDM physics, and as a means to characterize how DES observables provide cosmological information. We analyze the same multi-probe DES data as Ref.~\cite{Abbott:2018wzc}: DES Year 1 (Y1) galaxy clustering and weak lensing, which are sensitive to both growth and geometry, as well as Y1 BAO and Y3 supernovae, which probe geometry. We additionally include external geometric information from BOSS DR12 BAO and a compressed Planck 2015 likelihood, and external growth information from BOSS DR12 RSD. We find no significant disagreement with $\Omega_m^{\rm grow}=\Omega_m^{\rm geo}$. When DES and external data are analyzed separately, degeneracies with neutrino mass and intrinsic alignments limit our ability to measure $\Omega_m^{\rm grow}$, but combining DES with external data allows us to constrain both growth and geometric quantities. We also consider a parameterization where we split both $\Omega_m$ and $w$, but find that even our most constraining data combination is unable to separately constrain $\Omega_m^{\rm grow}$ and $w^{\rm grow}$. Relative to $\Lambda$CDM, splitting growth and geometry weakens bounds on $\sigma_8$ but does not alter constraints on $h$.
\end{abstract}

\preprint{DES-2019-0520}
\preprint{FERMILAB-PUB-20-529-AE}

\maketitle

\section{Introduction}\label{sec:intro}

One of the major goals of modern cosmology is to better understand the nature of the dark energy that drives the Universe's accelerating expansion. Though the simplest model for dark energy, a cosmological constant $\Lambda$, is in agreement with nearly all observations to date, there  exist a number of viable alternative models which explain the observed acceleration by introducing new fields or by extending general relativity via some form of modified gravity~\cite{Joyce:2014kja,Weinberg:2012es}.
Because there is no single most favored theoretical alternative, observational studies of dark energy largely consist of searches for tensions with the predictions of a minimal cosmological model, \lcdm, which consists of  a cosmological constant description of dark energy ($\Lambda$), cold dark matter (CDM), and general relativity as the theory of gravity.

A tension that has attracted significant  attention is one between constraints on the amplitude of matter density fluctuations $\sigeight$ made by  low redshift measurements, e.g. by the Dark Energy Survey (DES), and by Planck measurements of the Cosmic Microwave Background (CMB).
This comparison is often phrased in terms of $S_8\equiv \sigeight\sqrt{\Omega_m/0.3}$, the parameter combination most constrained by weak lensing analyses.
Though the DES and Planck results are not in tension according to the statistical metrics used in the original DES Year 1 analysis~\cite{Abbott:2017wau} (note that this is a topic of some discussion~\cite{Handley:2019wlz}),  the DES constraints prefer slightly lower \sigeight than those from Planck. This offset is in a direction consistent with other lensing results~\cite{Hildebrandt:2016iqg,Joudaki:2016mvz,Joudaki:2017zhq,Leauthaud:2016jdb,vanUitert:2017ieu,Joudaki:2017zdt,Hikage:2018qbn,Hamana:2019etx,Heymans:2020gsg}, and has been demonstrated to be independent~\cite{Park:2019tyw} of the much-discussed tension between CMB and local SNe measurements of the Hubble constant $H_0$~\cite{Riess:2016jrr,Riess:2019cxk,Knox:2019rjx}.
In fact, of the
numerous theoretical studies focused on alleviating the $H_0$ tension, most have found
 a joint resolution of the $\sigeight$ and $H_0$ tensions challenging, as discussed in e.g.~Refs.~\cite{Hill:2020osr,DiValentino:2017oaw,Clark:2020miy,Chen:2020iwm}.
Independent CMB measurements from ACT and WMAP give  \sigeight constraints  consistent with those from Planck~\cite{Aiola:2020azj}, while constraints based solely on reconstructed Planck CMB lensing maps are consistent with \sigeight constraints from both DES and measurements of CMB temperature and polarization~\citep{DESSPT:2019,Bianchini:2019vxp}.

These tensions are interesting because mismatched constraints from low and high-redshift probes  could indicate a need to extend our cosmological model beyond \lcdm.
Of course,  it is also possible that these offsets could be caused by
systematic errors or  a statistical fluke.  Given this, it is important to examine how different observables contribute to the \sigeight (and $H_0$) tension, as well as what classes of model extensions have the potential to alleviate them.

With this goal in mind, we perform a consistency test between geometric measurements of  expansion history and measurements of the growth of  large scale structure.
The  motivation for this test is similar to that of the early- vs. late-Universe (Planck vs. DES) comparison: we want to check for agreement between two classes of cosmological observables that have been split in a physically motivated way.
More ambitiously, we can also view this analysis as a search for signs of beyond-\lcdm physics. The  growth-geometry split is motivated in particular by the fact that modified gravity models  have been shown to generically break the consistency between expansion and structure growth expected in \lcdm \cite{Ishak:2005zs,Linder:2005in,Knox:2005rg,Bertschinger:2006aw,Huterer:2006mva}.

Our analysis focuses on data from the Dark Energy Survey (DES).  
DES is an imaging survey conducted between 2013-2019 which mapped galaxy positions and shapes over a 5000~deg$^2$ area and performed a supernova survey  in a smaller 27~deg$^2$ region.  
 This large survey volume and access to multiple observables make DES a powerful tool for constraining both expansion history and structure growth. 
  Constraints on cosmological parameters from the first year of DES data (Y1) have 
been published for the combined analysis of galaxy clustering and weak 
lensing~\cite{Abbott:2017wau,Abbott:2018xao}, for the baryonic acoustic oscillation (BAO) 
feature in the galaxy distribution~\cite{Abbott:2017wcz}, and for galaxy cluster abundance~\cite{Abbott:2020knk}. Additionally, 
cosmological results have been reported for the first three years (Y3) of 
supernova  data~\cite{Abbott:2018wog}, as well as for the combined analysis of Y3 SNe with Y1 galaxy clustering, weak lensing, and BAO~\cite{Abbott:2018wzc}. 
Analyses of  DES Y3 clustering and lensing data  are currently underway. The results presented in this paper are based on a multi-probe analysis like that of Ref.~\cite{Abbott:2018wzc}.

Because weak lensing and large scale structure probes like those measured by DES mix information from growth and geometry~\cite{Abazajian:2002ck,Jain:2003tba,Zhang:2003ii,Simpson:2004rz,Knox:2005rg,Zhan:2008jh}, rather than purely comparing \lcdm constraints from two datasets, we introduce new parameters to facilitate this comparison. As we explain in more detail in \sect{sec:modeling}, we define separate ``growth''  and ``geometry'' versions of a subset of cosmological parameters \params: \paramsgrow and \paramsgeo.
 By constraining growth and geometry parameters simultaneously, we can answer questions like
\begin{itemize}
\item Are DES constraints driven more by growth or geometric information?
\item Are the data consistent with the predictions of \lcdm --- that is, with $\paramsgrow=\paramsgeo$?
\item Is the DES preference for low \sigeight compared to Planck driven more by its sensitivity to background expansion (geometry) or by its measurement of the evolution of inhomogeneities (growth)?
  \end{itemize}
Our analysis thus serves as both a model-independent search for new physics affecting structure growth and an approach to building a deeper understanding of how DES observables contribute cosmological information.

The closest predecessors to the present work are Refs.~\cite{Wang:2007fsa,
  Ruiz:2014hma, Bernal:2015zom} which introduce similar growth-geometry
consistency tests and apply them to data. These analyses have the same general
idea and approach as the present analysis, but differ in several important aspects of how they implement the theoretical modeling of observables in their split parameterization. 
In a similar spirit, Ref.~\cite{Lin:2017ikq} explores growth-geometry consistency without introducing new parameters, using instead dataset comparisons in a search for discordance with \lcdm.
These approaches are complemented by other attempts at model independent tests of dark energy and modified gravity~\cite{Zhang:2007nk, Amendola:2012ky}, including analyses involving meta
parameters analogous to our split parameterization~\cite{Abate:2008au,Chu:2004qx, Matilla:2017rmu}, as well as other parameterizations which allow structure growth to deviate from expectations set by general relativity. 
These include analyses that have constrained free amplitudes  multiplying the growth rate \fsig~\cite{Alam:2016hwk}, or the ``growth
index'' parameter $\gamma$~\cite{Linder:2005in,Basilakos:2019hlb}. The commonly studied $\Sigma-\mu$ model of modified gravity~\cite{Ferte:2017bpf,Simpson:2012ra,Ade:2015rim,Joudaki:2016kym,Abbott:2018xao} is also in this category. In fact, the analysis presented below can be viewed as analogous to a $\Sigma$-$\mu$ study like that in Refs.~\cite{Garcia-Quintero:2020mja,Linder:2020xza}, with  $\Sigma$ fixed to its GR value, though differences  in our physical interpretation of the added parameters changes how we approach  analysis choices related to nonlinear scales.

\subsection{Plan of analysis}\label{sec:plan}

Our goal  is to test the consistency between DES Year 1 constraints from expansion  and those from measurements of the growth of large scale structure.
We will do this using using three different combinations of data:
\begin{enumerate}
  \item DES data
alone (including DES galaxy clustering and weak lensing, BAO, and supernova
measurements) --- henceforth, ``DES-only'' or just ``DES'';
\item As above, plus external data constraining geometry only from Planck 2015 and
  BOSS DR12 BAO measurements --- henceforth, ``DES+Ext-geo'';
\item As above, plus external growth information from BOSS DR12
  RSD measurements --- henceforth, ``DES+Ext-all''.
\end{enumerate}
Our main results will come from the combination of all of these datasets, but we will use the DES-only and DES+Ext-geo subsets to aid our interpretation of how different probes contribute information.\footnote{We do not include constraints from Planck 2018~\cite{Aghanim:2018eyx}, eBOSS DR14~\cite{Ata:2017dya,Agathe:2019vsu,Blomqvist:2019rah}, or eBOSS DR16~\cite{Alam:2020sor}  because those likelihoods were not available when we set up this analysis. At the end of this paper, in \sect{sec:outlook}, we will briefly discuss how updating to use those datasets might influence our results.}

The motivation for this growth-geometry split parameterization is to
study the mechanism behind late-time acceleration, so we focus on splitting
parameters associated with dark energy properties. Primarily, we will focus on the case  where we split
the matter density parameter \om in flat \lcdm,
that is
\lneqb\begin{equation}
  \om \rightarrow\{\omgeo, \omgrow\} \qquad\mbox{[Split \om]}.
\nonumber
\end{equation}\lneqe
As we discuss in more detail below, with some caveats, this split essentially means that $\omgeo$ controls quantities like 
comoving and angular distances, while $\omgrow$ controls quantities like the growth factor.  Because we impose the relation $\om + \Omega_{\Lambda}=1$, this means we also split $\Omega_{\Lambda}$, and $\omgrow\neq\omgeo$ necessarily implies $\Omega_{\Lambda}^{\rm growth} \neq \Omega_{\Lambda}^{\rm geo}$.

We will additionally show limited \wcdm results
where we split both \om and the dark energy equation of state, $w$, that is
\lneqb\begin{equation}
  \{\om, w\} \rightarrow
  \{\omgeo, \omgrow\, \wgeo, \wgrow\} \quad\mbox{[Split \om, $w$]}.
  \nonumber
\end{equation}\lneqe
Similarly to the split \om case, $\wgeo$ enters into calculations of comoving distances, while $\wgrow$ is used to compute, e.g., the growth factor.
We wish to calculate the posteriors for the split parameters given the aforementioned data, and in particular test their consistency (whether $\paramsgeo=\paramsgrow$) and identify any tensions.

For the split \om model  we will additionally examine how fitting in the
extended growth-geometry split parameter space   affects constraints on other
parameters, with an eye toward understanding degeneracies between the
split parameters and \mnu, \sigeight,
$ h\equiv H_0/\left[100\text{ km}\,\text{s}^{-1}\,\text{Mpc}^{-1}\right]$, and \aia. This will allow us to
build a deeper understanding of how the various datasets we consider provide
growth and geometry information. It will also allow us to weigh in on whether non-standard cosmological structure growth could potentially alleviate tensions between late- and early-Universe measurements of \sigeight and $h$.

Unless otherwise noted,  we use the same modeling and analysis choices as the DES Year 1 cosmology analyses described in Refs.~\cite{Krause:2017ekm, Abbott:2018xao, Abbott:2018wzc}. In order to ensure that our results are robust against various modeling choices and priors, we will follow similar blinding and validation procedures to those used in Ref.~\cite{Abbott:2018xao}'s analyses of DES Y1 constraints on beyond-\wcdm physics.

The paper is organized as follows.
In \sect{sec:modeling} we describe how we model observables in our growth-geometry
split parameterization, and in
\sect{sec:data} we introduce the data used to measure those observables.  \sect{sec:analysis} discusses our analysis
procedure, including the steps taken to protect our results from confirmation bias in \sect{sec:validation}, and our approach to quantifying tensions and model comparison in  \sect{sec:postunbl}. We present our main results, which are constraints on the split parameters and their consistency with \lcdm, in \sect{sec:results}. \sect{sec:bigplot_results} contains additional results characterizing how our growth-geometry split parameterization impacts constraints on other cosmological parameters, including  $\sigeight$. We conclude in \sect{sec:conclusion}. We discuss validation tests in detail in Appendices~\ref{sec:varyzi}-\ref{app:realdattests}, and in \app{sec:moreplots} we show plots of results supplementing those in the main body of the text.

\section{Modeling growth and geometry}\label{sec:modeling}

We consider several cosmological observables in our analysis: galaxy
clustering and lensing, BAO, RSD, supernovae and the CMB power
spectra.  We model these observables in a way that explicitly
separates information from geometry (i.e. expansion history) and
growth.  The separation of growth and geometry is immediately clear
for some probes; supernovae, for instance, are purely geometric
because they directly probe the luminosity distance.  For other
probes, however, this split is not obvious, or even necessarily
unique.  Throughout, we endeavor to make physically motivated,
self-consistent choices, and will note where past studies of growth
and geometry differ.  We emphasize that we are not developing a new
physical model, but are rather developing a phenomenological split of
$\Lambda$CDM.

Since one of our primary interests is in probing the physics associated with cosmic acceleration, we will use ``growth'' to describe the evolution of density perturbations in the late Universe.   Below, we describe our approach to modeling the observables we consider, and summarize this information in \tab{tab:modeling}.

Because structure growth depends primarily on the matter density via $\rho_m\propto h^2\om$ and we would like to decouple this from expansion-based constraints on $h$, for both our split parameterizations we additionally split the dimensionless Hubble parameter $h\equiv H_0/(100\, \text{km}\,\text{s}^{-1}\,\text{Mpc}^{-1})$. In practice we fix $h^{\rm grow}$ to a fiducial value because it has almost no effect on growth observables: varying $h$ across its full prior range results in fractional changes that are less than a percent for all observables considered. We demonstrate in \app{app:realdattests} that altering this choice by either not splitting $h$ or marginalizing over $h^{\rm grow}$ has little impact on our results.

{ \setlength{\tabcolsep}{12pt}
\begin{table*}
  \begin{center}
    \caption{Modeling summary. }
    \label{tab:modeling}
      \begin{tabular}{l l c  c c }
      \textbf{Observable}&\textbf{Modeling Ingredient} & \textbf{Described in} & \textbf{Geometry} & \textbf{Growth} \\\hline
      Galaxy clustering and lensing 
      & $P(k)$ shape at $z_i$ & \sect{sec:powerspec} & \checkmark &   \\
      & $P(k)$ evolution since $z_i$ & \sect{sec:powerspec} & & \checkmark   \\
      & Projection to 2PCF &  \sect{sec:modeling-mpp} & \checkmark &   \\
      & Intrinsic alignments & \sect{sec:modeling-mpp} & & \checkmark   \\\hline
      BAO & Distances &\sect{sec:modeling-bao} & \checkmark & \\\hline
      RSD 
      & $f(z)\sigeight(z)/\sigeight(0)$ &\sect{sec:modeling-rsd} & & \checkmark \\
      & $\sigeight(z=0)$ &\sect{sec:modeling-rsd} &\checkmark  & \checkmark  \\\hline
      Supernovae (SN)& Distances  & \sect{sec:modeling-sn} & \checkmark & \\\hline
      CMB & Compressed likelihood & \sect{sec:modeling-cmb} & \checkmark& \\ \hline
    \end{tabular}
  \end{center}

\end{table*}
}

\subsection{Splitting the matter power spectrum}\label{sec:powerspec}

Several of the observables that we consider depend on the matter power spectrum, namely galaxy clustering and lensing, RSD, and the CMB power spectrum. The matter power spectrum $P(k,z)$ contains both growth and geometric
information, so there is not a unique choice for how to compute it within our split parameterization. 
We choose a simple-to-implement and physically
motivated  approach.  Because we use ``growth''  to  describe the evolution of perturbations in the late Universe, we assume that  the early-time shape of the power spectrum is
determined by geometric parameters.  

More concretely, we construct the split linear power spectrum as a function of wavenumber $k$ and redshift $z$, $\plin^{\rm split}(k,z)$, by combining linear matter power
spectra computed separately using  geometric or growth parameters:
\lneqb\begin{equation}\label{eq:mixpower}
  \plin^{\rm split}(k,z) \equiv \frac{\plingeo(k,z_i)}{\plingrow(k,z_i)} \plingrow(k,z),
\end{equation}\lneqe
where $\plingeo$ and $\plingrow$ are the linear matter power spectra computed in $\Lambda$CDM using the geometric and growth parameters, respectively, and $z_i$ is an arbitrary redshift choice, to be discussed below. 
This definition has several desirable properties.  First, if the growth and geometric parameters are the same, then it reduces to the standard $\Lambda$CDM linear power spectrum.  Second, ignoring scale-dependent growth from neutrinos, $\plingrow(k,z)/\plingrow(k,z_i) = D^2(z)/D^2(z_i)$, where $D(z)$ is the linear growth factor.  Consequently, the growth parameters will effectively control the growth of perturbations from $z_i$ to $z$.  Third, for $z \ll z_i$, this ratio of growth factors approaches one, so the early time matter power spectrum is controlled by the geometric parameters, as desired.

We compute nonlinear corrections to the matter power spectrum using
\halofit~\cite{Smith:2002dz,Bird:2011rb,Takahashi:2012em}. \halofit provides a recipe, calibrated on simulations, for converting the linear matter power spectrum into the nonlinear power spectrum. As arguments to the
\halofit fitting function, we use the mixed linear power spectrum from
\eq{eq:mixpower}, and use the growth versions of the cosmological
parameters. By using the growth parameters as arguments to \halofit, we ensure that nonlinear evolution is controlled by the growth parameters, and that if $\paramsgrow=\paramsgeo$, the resultant power spectrum agrees with that computed in the standard DES analyses of e.g., Ref.~\cite{Abbott:2017wau}.
Although \halofit has not been explicitly validated for our growth-geometry split model,
 using it is reasonable because  we are performing a consistency test against \lcdm rather than implementing a real physical model. 

\begin{figure}
  \centering
\includegraphics[width=\linewidth]{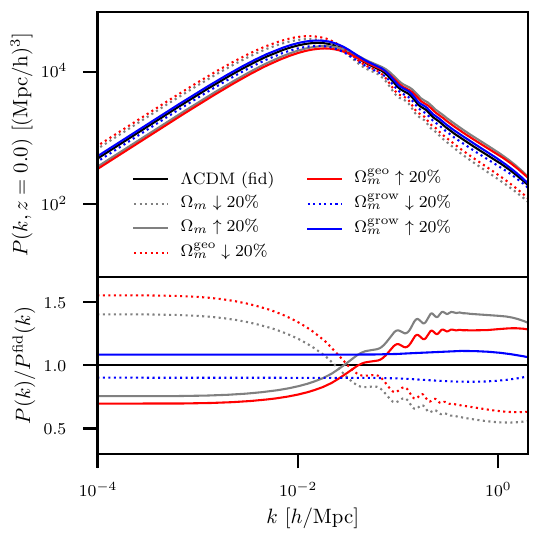}
  \caption{Dependence of the  nonlinear matter power spectrum on
    \omgrow and \omgeo. Gray lines show the impact of changing \om by
    $\pm20\%$ in \lcdm, red lines show changes to \omgeo, and blue lines show
    changes to \omgrow. The fiducial model uses $\omgrow=\omgeo=0.295$. Solid
    lines correspond to an increase in the relevant \om parameter to 0.354,
    while dotted lines show a decrease to 0.236. }
  \label{fig:PkvaryOm}
\end{figure}

\fig{fig:PkvaryOm} shows how the full nonlinear power spectrum
  $P(k,z=0)$ is affected by 20\% changes to \omgrow (blue) and \omgeo
  (red). For comparison, we also show the effect of changes to \om in
  \lcdm (gray).  The main effects of changing \omgeo are a scaling of the 
  normalization of the  power spectrum and a change in 
  the wavenumber where it peaks.
  This amplitude change occurs because the Poisson equation relates gravitational potential fluctuations $\Phi$ to matter density fluctuations $\delta$ via 
  \lneqb
  \begin{equation}
    k^2\,\Phi(k,z) =4\pi G \rho_m \delta = \tfrac{3}{2} H_0^2\om (1+z)\, \delta(k,z).
    \end{equation}
  \lneqe
  Thus, for fixed primordial potential power spectrum, the  matter power spectrum's early-time amplitude is proportional to $(\omgeo)^{-2}$. 
  The peak of the power spectrum occurs at the wavenumber corresponding to the horizon scale at matter-radiation equality, $k_{\rm eq} \propto \om h^2$, so increasing \omgeo shifts the peak to higher $k$. 
  Thus, the net effect of increasing \omgeo is a decrease
  in power at low $k$ and an increase in power at high $k$.  Changing
  \omgrow, on the other hand, impacts the late time growth, leading to
  a roughly scale-independent change in the power spectrum.  Nonlinear
  evolution at small scales breaks this scale indepdence.

We use $z_i=3.5$ as our fiducial value for the redshift at which growth
parameters start controlling the evolution of the matter power spectrum. This
choice is motivated by the fact that $z=3.5$ is before the dark energy
dominated era and is well beyond the redshift range probed by the DES samples.
Raising $z_i$ will slightly  increase the sensitivity to growth because it
 means that the growth parameters control a greater portion of the history of structure
  growth between recombination and the present. 
  However, as
long as $z_i$ is high enough, this has only a small effect on observables. For
the values of \omgrow and \omgeo shown in \fig{fig:PkvaryOm}, we confirm that
increasing $z_i$ to 5 or 10 results in changes of less than one percent at all
wave numbers of $\plin^{\rm split}(k,z=0)$, and also at all angular scales of the DES galaxy clustering and weak lensing 2pt functions.
Therefore, the combined constraints of DES and external data are weakly sensitive to the choice of $z_i$ as we show in \app{sec:varyzi}.

\subsection{Weak lensing and galaxy clustering}\label{sec:modeling-mpp}

For a photometric survey like DES, galaxy and weak lensing
correlations are typically measured via angular two-point correlation functions (2PCF).
To make theory predictions for 2PCF we first compute the angular power spectra.
Assuming flat geometry and using the Limber approximation~\cite{Limber:1953,LoVerde:2008re},
the angular power  spectrum between the $i$th redshift bin of tracer $A$ and the $j$th redshift bin of tracer $B$ is
\lneqb\begin{equation}\label{eq:cldef}
  C_{A B}^{ij}(\ell) = \left.\int \,dz\frac{H(z)}{c\,\chi^2(z)}W^i_{A}(z)W^j_{B}(z)P(k,z)\right|_{k=\left(\ell+\tfrac{1}{2}\right)/\chi(z)}.
\end{equation}\lneqe
Here $\chi$ is the comoving radial distance and $H(z)/(c\,\chi^2(z))$ is a
volume element that translates three-dimensional density fluctuations into
two-dimensional projected number density per redshift. The terms $W^i_{A}$ and
$W^j_{B}$ are window functions relating fluctuations in tracers $A$ and $B$
to the underlying matter density fluctuations whose statistics are described
by the power spectrum $P(k,z)$. The window functions for galaxy number density
$g$ and weak lensing convergence $\kappa$ are, respectively
\begin{align}
  W^i_g(z,k) =&\, n_i(z)\,b_i(z,k),\label{eq:Wg}\\
  W^i_{\kappa}(z) =& \left(\frac{3H_0^2\Omega_{m}}{2c^2}\right) \left(\frac{\chi(z)}{a(z)\,H(z)}\right)\nn
  &\times \int_z^{\infty}dz'n_i(z')\frac{\chi(z') - \chi(z)}{\chi(z')}.\label{eq:Wkappa}
\end{align}
In these expressions, $n_i(z)$ is the normalized redshift distribution of
galaxies in  sample $i$ while $b_i(z,k)$ is their galaxy bias. Following the
DES Y1 key-paper analysis \cite{Abbott:2017wau}, we will assume a constant linear bias for each sample, denoted with the parameter $b_i$.

In our growth-geometry split framework, we compute the power spectrum $P(k,z)$ in \eq{eq:cldef} via the procedure described in \sect{sec:powerspec}. We treat all projection operations in \eqs{eq:cldef}-(\ref{eq:gammat}) as geometric. This choice means that the usual \sigeight-\om weak lensing degeneracy will occur between between \sigeight (computed with $\plin^{\rm split}$) and \omgeo, so we  define $S_8\equiv \sigeight\sqrt{\omgeo/0.3}$.

We include contributions to galaxy shear correlations from intrinsic alignments between galaxy shapes via a non-linear alignment model~\cite{Bridle:2007ft} which  is the same intrinsic alignment model used in previous DES Y1 analyses~\cite{Krause:2017ekm}.  This model adds a term to the shear convergence window function,
\lneqb\begin{equation}\label{eq:iadef}
    W^i_{\kappa}(z) \rightarrow W^i_{\kappa}(z) -\aia\left[ \left(\frac{1+z}{1+z_0}\right)^{\alpha_{IA}}\frac{C_1\rho_{m0}}{D(z)}\right]n_i(z).
\end{equation}\lneqe
 Here \aia and $\alpha_{IA}$ are free parameters which
should be marginalized over when performing parameter inference.  The
normalization $C_1=0.0134/\rho_{\rm crit}$ is a constant calibrated based
on SuperCOSMOS observations~\cite{Bridle:2007ft}, $\rho_{m0}$ is the present-day physical matter
density, and $D(z)$ is the linear growth factor. Because intrinsic alignments
are caused by cosmic structures, in our split formulation, we compute these quantities using growth parameters.

To obtain real-space angular correlation functions which can be compared to DES measurements, we then transform the angular power spectra of \eq{eq:cldef} using Legendre and Hankel transformations. The correlation between galaxy positions in tomographic bins $i$ and $j$ is
\lneqb\begin{equation}
  w^{ij}(\theta) = \sum_{\ell}\frac{2\ell+1}{4\pi}P_{\ell}(\cos{\theta})\,C_{gg}^{ij}(\ell),
\end{equation}\lneqe
where $P_{\ell}(x)$ is the Legendre polynomial of order $\ell$.
Shear correlations
are computed in the flat-sky approximation as
\begin{align}
  \begin{split}
  \xi_+^{ij}(\theta) &= \int\frac{d\ell\, \ell}{2\pi}J_0(\ell\theta)\, C_{\kappa \kappa}^{ij}(\ell),\\
  \xi_-^{ij}(\theta) &= \int\frac{d\ell\, \ell}{2\pi}J_4(\ell\theta)\, C_{\kappa \kappa}^{ij}(\ell).
  \end{split}
\end{align}
In these expressions, $J_m(x)$ is a Bessel function of the first kind of order
$m$. Finally, the correlation between galaxy positions in bin $i$ and tangential shears
in bin $j$ --- the so-called ``galaxy-galaxy lensing'' signal --- is similarly computed via
\begin{align}
  \gamma_t^{ij}(\theta) &= \int\frac{d\ell \,\ell}{2\pi}J_2(\ell\theta)\, C_{g \kappa}^{ij}(\ell).\label{eq:gammat}
\end{align}
In our analysis, we perform these Fourier transformations using the function {\tt tpstat\_via\_hankel} from the {\sc nicaea} software.\footnote{\url{www.cosmostat.org/software/nicaea}}~\cite{Kilbinger:2008gk}

Several astrophysical and measurement systematics impact observed correlations for galaxy clustering and weak lensing. In addition to intrinsic alignments, which we addressed above, these include shear calibration and photometric redshift uncertainties.
We model these effects following the previously published DES Y1 analyses~\cite{Krause:2017ekm}, introducing several nuisance parameters that we marginalize over when performing parameter estimation. This includes a shear calibration parameters $m_i$ for each redshift bin $i$ where shear is measured and a photometric redshift bias parameter $\Delta z_i$ for each redshift bin $i$.  These systematic effects are not cosmology dependent and so are not impacted by the growth-geometry split.

\subsection{BAO}\label{sec:modeling-bao}
Baryon acoustic oscillations (BAO) rely on a characteristic scale imprinted on galaxy clustering which is set by the sound horizon scale at the end of the Compton drag epoch. That characteristic physical sound horizon scale is
\lneqb\begin{equation}
  r_d = \int_{z_d}^{\infty}\frac{c_s(z)}{H(z)}dz,
\end{equation}\lneqe
where $c_s$ is the speed of sound, $z_d$ is the redshift of drag epoch, and $H(z)$ is the expansion rate at redshift $z$.
Measurements of the BAO feature in galaxy clustering in directions transverse
to the line of sight constrain $D_M(z)/r_d$, where $D_M(z)=(1+z)D_A(z)$ is the
comoving angular diameter distance and $D_A(z)$ is the physical angular
diameter distance. Line-of-sight  measurements, on the other hand, constrain  $H(z)r_d$.
In practice constraints from BAO analyses are reported in terms of  dimensionless ratios,
\lneqb\begin{equation}\label{eq:alphaperp}
  \alpha_{\perp} = \frac{D_M(z)r_{d}^{\rm fid}}{D_M^{\rm fid}(z)r_{d}}
\end{equation}\lneqe
and
\lneqb\begin{equation}\label{eq:alphapar}
  \alpha_{\parallel} = \frac{H^{\rm fid}(z)r_{d}^{\rm fid}}{H(z)r_{d}},
\end{equation}\lneqe
where the superscript ``fid'' indicates that the quantity is computed at a fiducial cosmology.

The
cosmological information here comes fundamentally from measures of distances via the comparison between the observed scale of the BAO feature and the physical distance $r_d$.  Given this, in our split parameterization we compute the expressions in \eqs{eq:alphaperp} and~(\ref{eq:alphapar}) using geometric parameters.

\subsection{RSD}\label{sec:modeling-rsd}

Redshift-space distortions (RSD) measure anisotropies in the apparent
clustering of matter in redshift space. These distortions are caused by the
infall of matter into overdensities, so the RSD allow us to measure the rate
of growth of cosmic structure.
RSD constraints are presented in terms of constraints on $f(z)\sigma_8(z)$,
where $f(z)=d\ln D/d\ln a$ for linear density fluctuation amplitude $D$ and
scale factor $a = (1+z)^{-1}$. In   our split parameterization,  the amplitude $\sigeight\equiv \sigeight(z=0)$ should match the value computed using the  mixed power spectrum $\plin^{\rm split}$ from \eq{eq:mixpower}, while the time evolution of $\sigeight(z)/\sigeight(0)$ and the growth rate $f(z)$ should be governed by growth parameters.

To achieve this, we proceed as follows.  First, following the method used in
Planck analyses~\cite{Ade:2015xua} (see their Eq.~33), we use our growth parameters to compute
\lneqb\begin{equation}\label{eq:fsigeight}
  f(z)\sigeight^{\rm grow}(z) = \frac{\left[\sigeight^{(\delta v)}(z)\right]^2}{\sigeight^{\rm grow}(z)}.
\end{equation}\lneqe
Here the  superscript on $\sigeight^{\rm grow}$ denotes that it was computed within \lcdm
using the growth parameters.  The quantity $\sigma^{(\delta v)}_8$ is the
smoothed density-velocity correlation; it is defined similarly to
$\sigeight(z)$, but instead of using the matter power spectrum $P(k,z)$ it is
computed by integrating over the linear cross power between the matter density
fluctuations $\delta$  and the divergence of the dark matter and baryon (but
not neutrino) peculiar velocity fields in Newtonian-gauge, $v=-\nabla \vec{v}_N/H$. Ref.~\cite{Ade:2015xua} motivates this definition by noting that it is close to what is actually being probed by RSD measurements.

In order to make \sigeight consistent with our split matter power spectrum
definition from \eq{eq:mixpower}, we multiply \eq{eq:fsigeight}  by the
$z=0$ ratio of \sigeight, computed from $\plin^{\rm split}$, and
$\sigeight^{\rm grow}$. The quantity that we use to compare theory
  with RSD measurements is therefore:
\lneqb\begin{equation}
  f(z)\sigeight(z) = f(z)\sigeight^{\rm grow}(z)\times\frac{\sigeight(0)}{\sigeight^{\rm grow}(0)}.
\end{equation}\lneqe
This expression will be consistent with our method of defining the linear power spectrum in \eq{eq:mixpower} as long as it is evaluated at $z<z_i$.

\subsection{Supernovae}\label{sec:modeling-sn}

Cosmological information from supernovae  comes from measurements of the
apparent magnitude of Type Ia supernovae as a function of redshift.
Because
the absolute luminosity of Type Ia supernovae can be calibrated to serve as
standard candles,
the observed flux can be used as a distance measure. Even when the value of that absolute luminosity is not calibrated with more local distance measurements, 
the relationship between observed supernova fluxes and redshifts contains information about how the expansion rate of the Universe has
changed over time.

Measurements and model predictions for supernovae are  compared
in terms of the distance modulus $\mu$, which is related to
the luminosity distance $d_L$ via
\lneqb\begin{equation}
  \mu = 5\,\log\left[d_L/10{\text pc}\right].
\end{equation}\lneqe
The observed distance modulus is nominally given by the sum of the apparent
magnitude, $m_B$, and a term accounting for the combination of the absolute
magnitude and the Hubble constant, $M_0$.

We follow the approach to computing this used in the DES Y3 supernovae analysis~\cite{Brout:2018jch}, 
also described in Ref.~\cite{Scolnic:2017caz}, and use the  \cosmosis module associated with the latter paper to perform the calculations. 
In practice, computing the distance requires a few additional modeling components. These include the width $x_1$ and color $\mathcal{C}$ of the
light curve, which are used to standardize the luminosity of the Type Ia supernovae, as well as a parameter $G_{\text{host}}$ which introduces a step function  to account for correlations between supernova luminosity and host
galaxy stellar mass $M_{\text{host}}$   ($G_{\text{host}}$  is $+1/2$ if $M_{\text{host}}>10^{10}M_{\odot}$, $-1/2$ if $M_{\text{host}}<10^{10}M_{\odot}$).
 The final expression for the distance modulus in terms of these
parameters is
\lneqb\begin{equation}\label{eq:snmu}
\mu = m_B + \alpha x_1 - \beta\mathcal{C} + M_0 + \gamma G_{\text{host}} + \Delta\mu_{\text{bias}}.
\end{equation}\lneqe
Here the calibration parameters $\alpha$, $\beta$, and $\gamma$  are fit to data using the formalism from Ref.~\cite{Marriner:2011mf},  and the selection bias  $\Delta\mu_{\rm bias}$ is calibrated using simulations~\cite{Kessler:2016uwi}. 
The parameter $M_0$ is marginalized over during parameter estimation.

The cosmological information in supernova observations comes from distance measurements, so in our split parameterization we compute these quantities using geometric parameters.

\subsection{CMB}\label{sec:modeling-cmb}

The cosmic microwave background (CMB) anisotropies in temperature and polarization are a rich cosmological observable with information about both growth and geometry. The geometric information primarily  consists of the distance to the last scattering surface and the sound horizon size at recombination. 
Two parameters encapsulate how these distances (and through them, the cosmological parameters) impact the observed CMB power spectra:
the shift parameter~\cite{Efstathiou:1998xx}, 
\lneqb\begin{equation}
  R_{\rm shift}\equiv\sqrt{\Omega_m(100h)^2} D_M(z_*)/c,
\end{equation}\lneqe
which describes the location of the first power spectrum peak, and the angular scale of the sound horizon at last scattering $\ell_A=\pi/\theta_*$,
\lneqb\begin{equation}
  \ell_A\equiv \pi D_M(z_*)/r_s(z_*).
\end{equation}\lneqe
Here $z_*$ is the redshift of recombination, $D_M$ is the comoving angular diameter distance at that redshift, and $r_s$ is the comoving sound horizon size. In our split parameterization, we use geometric parameters to compute these quantities.

The CMB is sensitive to late-time structure growth in a few different ways.  The ISW effect adds TT power at low-$\ell$ in a way that depends on the linear growth rate,  and weak lensing from low-$z$ structure smooths the peaks of the CMB power spectra at high-$\ell$. To be self-consistent, the calculation of these effects should use the split power spectrum described in \sect{sec:powerspec}. Adapting the ISW and CMB lensing predictions to our split parameterization would therefore require a modification of the {\sc CAMB} software\footnote{\url{http://camb.info}}~\cite{Lewis:1999bs,Howlett:2012mh} we use to compute power spectra. In order to simplify our analysis,  we focus on a subset of  measurements from the CMB that are closely tied to geometric observables, independent of late-time growth.

We do this via a compressed likelihood  which
describes CMB constraints on $R_{\rm shift}$, $\ell_A$, $\Omega_bh^2$, $n_s$, and \as after marginalizing over all other parameters, including \mnu and $A_{\rm Lens}$. This approach is inspired by the fact that the  CMB mainly probes expansion history, and thus dark energy, via the geometric information provided by the locations of its acoustic peaks \cite{Frieman:2002wi}, and by the compressed Planck likelihood provided in Ref.~\cite{Ade:2015rim}; see \sect{sec:cplanck} below for details. In this formulation, we have constructed our CMB observables to be independent of late-time growth,  so we compute the model predictions for them with geometric parameters.

\subsection{Modeling summary and comparison to previous work}\label{ggsplit-summary}

\tab{tab:modeling} summarizes the sensitivity of the  probes discussed above to growth and geometry. Briefly,  we derive constraints from structure growth from the LSS  observables --- galaxy clustering, galaxy-galaxy lensing, weak lensing shear, and RSD --- while all probes we consider provide some information about geometry. Constraints from BAO, supernovae, and the scale of the first peaks of the CMB provide purely geometric information. The LSS observables mix growth and geometry via their dependence on the power spectrum: its shape is set by geometry, while its evolution since $z_i=3.5$ is governed by growth parameters. All projection translating from three-dimensional matter power to two-dimensional observed correlations are geometry dependent.

We now compare our choices to previous work. 

For the CMB, our geometry-growth split choices are motivated by simple
implementation and (since our focus is on DES data) the ease of
interpretation. In this we roughly follow the approach in
Ref.~\cite{Ruiz:2014hma}, which also considers a compressed CMB likelihood that is
governed purely by geometry. In contrast, Ref.~\cite{Wang:2007fsa} describes
CMB fluctuations (and so the sound horizon scale) using growth parameters,
then uses geometry parameters in converting physical to angular scales.
Ref.~\cite{Bernal:2015zom} splits the growth and geometric information in the
CMB by multipole, using the TT, TE, and EE power spectra at $\ell>30$ to
constrain geometric parameters, and the low $\ell$ (<30) multipoles as well as
the lensing power spectrum to constrain growth. 

For weak lensing, our approach is
closest to Ref.~\cite{Ruiz:2014hma}, with an additional modification in how we
model the matter power spectrum, described in Sec.~\ref{sec:powerspec}. Ref.~\cite{Bernal:2015zom} leaves weak lensing out of their analysis, citing the difficulty in separating growth and geometric contributions to  those observables. Both Refs.~\cite{Wang:2007fsa} and~\cite{Ruiz:2014hma} compute the matter power spectrum entirely using growth parameters (as opposed to our split parameterization described in Sec.~\ref{sec:powerspec}) and (like us) they use geometric parameters for projection operations and for the distances used to compute the weak lensing kernel.   These analyses differ
in how they treat the lensing kernel's $\Omega_M^2$ prefactor  (see Eq.~(\ref{eq:Wkappa})).  Ref.~\cite{Wang:2007fsa} treats this as a growth quantity, while Ref.~\cite{Ruiz:2014hma} 
considers it part of the lensing window function and hence a geometric
quantity.  Our choice, which matches that of Ref.~\cite{Ruiz:2014hma}, means \omgrow affects weak lensing observables solely through changes in  the matter power spectrum. Though this weakens our ability to constrain \omgrow, it has the benefit of  making our model more phenomenologically similar to other parameterizations of non-standard structure growth, making the interpretation of results more easily generalizable. 

Our treatment of BAO and Type Ia supernovae agrees with all previous
literature in treating these probes as purely geometrical. Finally, our
treatment of the RSD is subtly different from previous literature on the
subject \citep{Wang:2007fsa, Ruiz:2014hma, Bernal:2015zom} which assumed
$f\sigma_8$ are determined purely by the growth parameters, Our RSD is
\textit{mostly} determined by the growth of structure, but we allow
$\sigma_8(z=0)$ to also include geometric parameters via our split
parameterization of the matter power spectrum.

\section{Data and likelihoods}\label{sec:data}
In this section we describe the data and likelihoods used for our analyses. The datasets and where to find their descriptions are summarized in \tab{tab:datasets}.

{ \setlength{\tabcolsep}{6pt}
\begin{table*}
  \begin{center}
    \caption{Table summarizing datasets included and abbreviations for plots.}
    \label{tab:datasets}
    \begin{tabular}{l l c c c c}

      \textbf{Combination} & \textbf{Datasets}&  \textbf{Described in}&\textbf{Geometry} & \textbf{Growth} \\\hline
      DES     & DES Y1 \mpp (galaxy clustering and WL)  &\sect{sec:3x2pt}& \checkmark & \checkmark \\
      & DES Y1 BAO &  \sect{sec:desbao} &\checkmark &   \\
      & DES Y3 + lowZ SNe & \sect{sec:dessn}& \checkmark & \\\hline
      Ext-geo & Compressed 2015 Planck likelihood  &\sect{sec:cplanck}& \checkmark & \\
      & BOSS DR12 BAO &\sect{sec:bossbao}& \checkmark & \\\hline
      Ext-all & Ext-geo  & & \checkmark &\\
      & BOSS DR12 RSD &\sect{sec:extgrow}& \checkmark &\checkmark \\

    \end{tabular}
  \end{center}
\end{table*}
}
\subsection{DES Year 1 combined data}\label{sec:desdata}

In our growth-geometry split analysis of DES data, we perform a combined analysis of DES Y1 galaxy clustering and weak lensing, DES Y1 BAO, and DES Y3 supernova measurements, following a similar methodology to the multi-probe analysis in Ref.~\cite{Abbott:2018wzc}. The combination of these measurements will be referred to as ``DES'' in the reported constraints below. We now describe the constituent measurements.

Galaxy samples used in these measurements were constructed from the DES Y1 Gold catalog~\cite{Drlica-Wagner:2017tkk}, which is derived from imaging data taken between August 2013 and February 2014 using the 570-megapixel Dark Energy Camera~\cite{Flaugher:2015pxc} at CTIO. The data in the catalog covers an area of 1321 deg$^2$ in grizY filters and were processed with the DES Data Management system~\cite{Desai:2012zr,Sevilla:2011ps,Mohr:2008tx,Morganson:2018zvt}.

\subsubsection{DES Y1 galaxy clustering and weak lensing}\label{sec:3x2pt}

\begin{figure}
  \centering
\includegraphics[width=\linewidth]{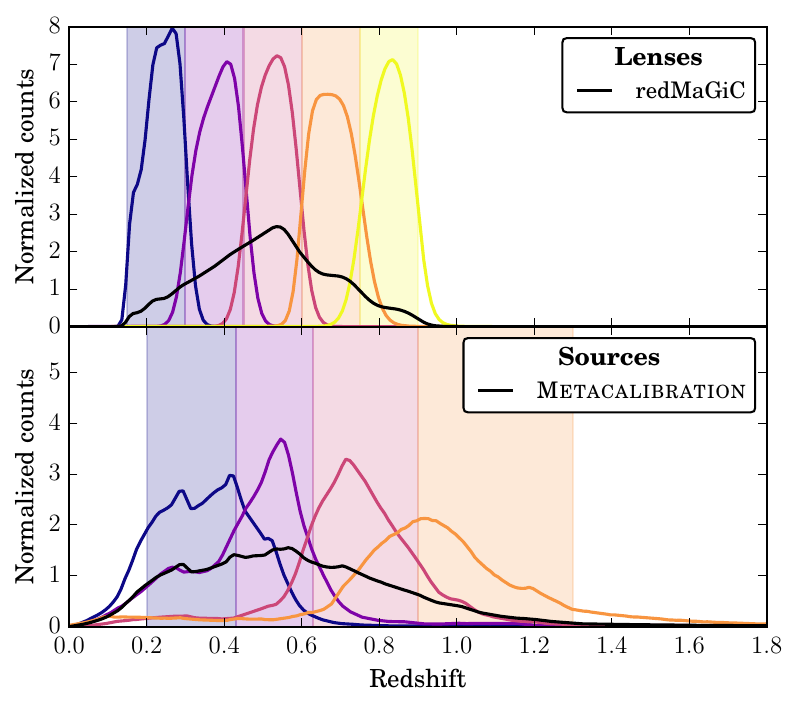}
  \caption{Redshift distribution of source and lens galaxies  used in the DES \yone analysis. The vertical shaded bands represent the nominal range of the redshift bins,  while the solid lines show their estimated true redshift distributions, given their photometric-redshift-based selection.}
  \label{fig:mpp_nz}
\end{figure}

The DES Y1 combined galaxy clustering and weak lensing analysis, hereafter \yone, is based on the analysis of three types of angular two-point correlation functions (2PCF): the correlation between the positions of a population of lens galaxies, between the measured shapes of a population of source galaxies, and the correlation of lens positions and source shapes. 
The lens galaxy sample consists of approximately 660,000 luminous red galaxies which were found using the \redmagic algorithm~\cite{Rozo:2015mmv} and were selected using luminosity cuts to have relatively small photo-$z$ errors. They are split into five redshift bins with nominal edges at $z=\{0.15,0.3,0.45,0.6,0.75,0.9\}$. 
Weak lensing shears are measured from the source galaxy sample, which includes 26 million  galaxies. These were selected from the Y1 Gold catalog using the \metacal~\cite{Huff:2017qxu,Sheldon:2017szh} and {\sc  NGMIX}\footnote{\url{https://github.com/esheldon/ngmix}} algorithms, and the {\sc BPZ} algorithm~\cite{Coe:2006hj} is used to estimate redshifts. The source galaxies are split into four redshift bins with approximately equal densities, with nominal edges at $z=\{0.2,0.43,0.63,0.9,1.3\}$~\cite{Zuntz:2017pso,Hoyle:2017mee}. For each source bin a multiplicative shear calibration parameter $m_i$ for $i\in\{1,2,3,4\}$ is introduced in order to prevent shear measurement noise and selection effects from biasing cosmological results. \metacal provides tight Gaussian priors on these parameters. The redshift distributions for the lens and source galaxies used in the DES Y1 galaxy clustering and weak lensing measurements are shown in \fig{fig:mpp_nz}. Uncertainties in photometric redshifts are quantified with nine nuisance parameters $\Delta z_i^x$ which quantify translations of each redshift bin's distribution to $n^x_i(z-\Delta z^x_i)$, where $i$ labels the redshift bin and $x=\text{source}$ or lens.

The 2PCF  measurements that comprise the \yone data are presented in Ref.~\cite{Elvin-Poole:2017xsf} (galaxy-galaxy), Ref.~\cite{Prat:2017goa} (galaxy-shear), and Ref.~\cite{Troxel:2017xyo} (shear-shear). Each 2PCF is measured in 20 logarithmic bins of angular separation from $2.5'$ to $250'$ using the {\sc Treecorr}~\cite{Jarvis:2003wq} algorithm. Angular scale cuts are chosen as described in Ref.~\cite{Krause:2017ekm} in order to remove measurements at small angular scales where our model is not expected to accurately describe the impact of nonlinear evolution of the matter power spectrum and baryonic feedback. The resulting DES \yone data vector contains 457 measured 2PCF values. The likelihood for the \mpp analysis is assumed to be Gaussian in that data vector. The covariance is computed using {\sc Cosmolike}~\cite{Krause:2016jvl}, which employs a halo-model-based calculation of four-point functions~\cite{Cooray:2002dia}. Refs.~\cite{Krause:2016jvl,Troxel:2018qll} present more information about the calculation and validation of the covariance matrix.

\subsubsection{DES Y1 BAO}\label{sec:desbao}

The measurement of the signature of baryon acoustic oscillations (BAO) in DES Y1 data is presented in Ref.~\cite{Abbott:2017wcz}.
That measurement is summarized as a likelihood of the ratio between the angular diameter distance and the drag scale $D_A(z=0.81)/r_d$. This result was derived from the analysis of a sample of 1.3 million galaxies from the DES Y1 Gold catalog known as the DES BAO sample.
These galaxies in the sample have photometric redshifts between 0.6 and 1.0 and were selected using color and magnitude cuts in order to optimize the high redshift BAO measurement, as is described in Ref.~\cite{Crocce:2017iwq}. An ensemble of  1800 simulations~\cite{Avila:2017nyy} and three different methods for measuring galaxy clustering~\cite{Ross:2017emc,Camacho:2018mel,Chan:2018pjr} were used to produce the DES BAO likelihood.

The DES BAO sample is measured from the same survey footprint as the samples used in the DES \yone analysis, so there will be some correlation between the two measurements. Following Ref.~\cite{Abbott:2018wzc}, we neglect this correlation when combining the two likelihoods. This can be motivated by the fact that the intersection between the \mpp and BAO galaxy samples is estimated to be about 14\% of the total BAO sample, and the fact that no significant BAO signal is measured in the 2PCF measured for the \mpp analysis.

\subsubsection{DES Y3 + lowZ Supernovae}\label{sec:dessn}

The cosmological analysis of supernova magnitudes from the first three years of DES observations is presented  in Ref.~\cite{Abbott:2018wog}. The 207 supernovae used in this analysis were discovered via repeated deep-field observations of in a 27 deg$^2$ region of the sky taken between  August 2013 and February 2016, and are in the redshift range $0.07<z<0.85$.
A series of papers describe the search and discovery~\cite{Morganson:2018zvt,Goldstein:2015cds,Kessler:2015mjg}, calibration~\cite{Burke:2017gns,Lasker:2018fnw}, photometry~\cite{Brout:2018ttj},  spectroscopic follow-up~\cite{DAndrea:2018sfi}, simulations~\cite{Kessler:2018krb}, selection effects~\cite{Scolnic:2016ukm}, and analysis methodology~\cite{Brout:2018jch} that went into those results.
Following the DES supernova analysis~\cite{Abbott:2018wzc,Abbott:2018wog} (but  not the fiducial choices of the multi-probe analysis of Ref.~\cite{Abbott:2018wzc}), we additionally include in the supernova sample the so-called low-$z$ subset: 122 supernovae at $z<0.1$ that were measured as part of the Harvard-Smithsonian Center for Astrophysics Surveys~\cite{Hicken:2009dk,Hicken:2012zr} and the Carnegie Supernova Project~\cite{Contreras:2009nt}.

The DES supernova likelihood is a multivariate Gaussian in the difference between the predicted and measured values of the distance modulus $\mu$. The likelihood is implemented in our analysis pipeline using the \cosmosis Pantheon~\cite{Scolnic:2017caz} module,  adapted to use the DES measurements instead of the original Pantheon supernova sample.

\subsection{External geometric data}\label{sec:extgeo}
\subsubsection{BOSS DR12 BAO}\label{sec:bossbao}
We use BAO information from the constraints presented in the BOSS  Data Release 12~\cite{Alam:2016hwk}. 
The likelihood provided by BOSS has a default fiducial $r_d$, and measurements on $D_M(z)$ and $H(z)$ (described in \sect{sec:modeling-bao}) at the  redshifts $z=\{0.38, 0.51, 0.61\}$.
These constraints include measurements of the Hubble parameter $H(z)$ and
comoving angular diameter distance $d_A(z)$ at redshifts $z=\{0.38, 0.51,
0.61\}$. Specifically, we use the post-reconstruction BAO-only consensus measurements data file
{\tt BAO\_consensus\_results\_dM\_Hz.txt} and covariance files {\tt
  BAO\_consensus\_covtot\_dM\_Hz.txt} provided on the BOSS results
page.\footnote{\label{foot:boss} \url{https://www.sdss3.org/science/boss\_publications.php}
} No
covariance with other data is assumed.

\subsubsection{Compressed Planck likelihood}\label{sec:cplanck}

In order to extract information from Planck data that is independent of our growth parameters, we make our own version of the compressed Planck likelihood presented in Ref.~\cite{Ade:2015rim}. This likelihood is a five-dimensional Gaussian likelihood extracted from a Multinest chain run with the Planck lite 2015 likelihood using the temperature power spectrum (TT) and low-$\ell$ temperature and polarization, with no lensing. We ran this chain using the same settings as used for the Planck constraints reported in the DES Y1 papers~\cite{Abbott:2017wau}, which includes fixing $w=-1$ and marginalizing over neutrino mass.
We also marginalize over the lensing amplitude $A_{\rm Lens}$ to reduce the possible impact of growth via weak lensing on the temperature power spectrum. From that chain we extracted a 5D mean and covariance for the parameter vector $[R_{\rm shift}, \ell_A, \Omega_bh^2, n_s, 10^9\as]$. The compressed likelihood is then a five-dimensional multivariate Gaussian in those parameters. We confirm that this compressed likelihood is an accurate representation of the Planck constraints in this five-dimensional parameter space --- in other words, that the Planck likelihood is approximately Gaussian ---  by checking that the chain samples for the full Planck likelihood follow a $\chi^2$ distribution when evaluated relative to the mean and covariance used in the compressed likelihood.

\subsection{External growth data (RSD)}\label{sec:extgrow}
We include an external growth probe using the BOSS DR12 combined results~\cite{Alam:2016hwk}. We use the full-power-spectrum-shape-based consensus measurements data file {\tt final\_consensus\_results\_dM\_Hz\_fsig.tx} and covariance file {\tt final\_consensus\_covtot\_dM\_Hz\_fsig.txt} provided on the BOSS results page.$^{\rm \ref{foot:boss}}$ This includes consensus measurements of $D_M(z)$, $H(z)$, and $f(z)\sigma_8(z)$ at the same three redshifts $z=\{0.38, 0.51, 0.61\} $ as the BAO-only likelihood. The reported values are the combined results from seven different measurements using different techniques and modeling assumptions, where the covariances between those results have been assessed using mock catalogues~\cite{Kitaura:2015uqa, Sanchez:2016gky}.

As a slight complication, we note that these BOSS results use both the post-reconstruction BAO-only fits described in \sect{sec:bossbao}, and those from the full-shape analysis of the pre-reconstruction data. The combination of the post-reconstruction BAO and pre-reconstruction full-shape fits tightens constraints on $D_M(z)$ by around 10\% and on $H(z)$ by 15-20\%. This means that in addition to adding growth information from RSD, our Ext-all data combination will also have slightly tighter geometric constraints than Ext-geo.

\section{Analysis choices and procedure}\label{sec:analysis}

We use the same parameters and parameter priors as previous DES Y1 analyses~\cite{Abbott:2017wau,Abbott:2018xao,Abbott:2018wzc}. For our split parameters, we use the same prior as their unsplit counterparts' priors in those previous analyses:
\begin{align}
  \om\text{, }\omgeo\text{, }\omgrow &\in [0.1, 0.9]\\
  w \text{, }\wgeo\text{, }\wgrow &\in [-2.0, 0.33]
\end{align}
We use the same  angular scale cuts for the DES Y1 weak lensing and LSS measurements, leaving
 457 data points in the weak lensing and galaxy clustering combined \mpp data
 vector. The DES BAO likelihood contributes another measurement (of $D_A(z=0.81)/r_d$), and the DES SNe likelihood is based on measurements of 329 supernovae (207 from DES, 122 from the low-$z$ sample). This means that our DES-only analysis is based on a total of 787 data points. The DES+Ext-geo analysis therefore has 798 data points (787 from DES, 5 from compressed Planck, 6 from BOSS BAO), and the DES+Ext-all analysis has 801 (same as DES+Ext-geo plus 3 BOSS RSD measurements).

 Calculations were done in the \cosmosis\footnote{
    \url{https://bitbucket.org/joezuntz/cosmosis/}} software package~\cite{Zuntz:2014csq}, using the
  same pipeline as the Y1KP, modulo changes to implement the growth-geometry
  split.
  For validation tests, chains were run with {\sc Multinest}
  sampler~\cite{Feroz:2007kg,Feroz:2008xx,Feroz:2013hea}, with low resolution
  fast settings of 250 live points, efficiency 0.3, and tolerance
  0.01. 
  For fits to data where we need both posteriors and Bayesian evidence, we use {\sc Polychord}~\cite{Handley:2015fda} with 250 live points, 30 repeats, and tolerance of 0.01. 
  Summary statistics and contour plots from chains are done using the {\sc GetDist}~\cite{Lewis:2019} software with a smoothing kernel of 0.5.

 As noted in \sect{sec:plan}, our main results will be products of parameter estimation and model comparison evaluated for
 \begin{itemize}
\item  Split \om  constrained with DES+Ext-geo, and
\item Split \om constrained with DES+Ext-all.
\end{itemize}
 This choice was based on simulated analyses performed before running parameter estimation on real data.  In these analyses we computed model predictions for observables at a fiducial cosmology, then analyzed those predictions as if they were measurements. By studying the relationship between the resulting posteriors  and the input parameter values we identified which model-data combinations are  constraining enough so that parameter estimates are unbiased by parameter-space projection effects. This is described in more detail in \app{sec:syndat}. For the DES+Ext-geo and DES+Ext-all constraints on split \om, we confirm that the input parameter values are contained within the 68\% confidence intervals of the synthetic-data versions of all marginalized posteriors plotted in this paper.

 We consider two additional sets of constraints: 
\begin{itemize}
\item Split \om  constrained by DES only, and
\item Split \om and $w$ constrained by DES+Ext-all.
\end{itemize}
Our simulated analyses revealed that the one-dimensional marginalized posteriors are impacted by significant projection effects. Given this, for these cases 
we do not report numerical parameter estimates or error bars, but we will still report model comparison statistics (to be discussed in \sect{sec:postunbl}) and show their two-dimensional confidence regions on plots. 
We do not consider constraints splitting both \om and $w$ for DES-only and DES+Ext-geo because these datasets are less constraining than DES+Ext-all and so are expected to suffer from even more severe projection effects.
A more detailed discussion of these projection effects and the parameter degeneracies which cause them can be found in \app{sec:syndat}.

We follow a procedure similar to that used in Ref.~\cite{Abbott:2018xao} to validate our analysis pipeline. Our goal is to characterize the robustness of our results to reasonable changes to analysis choices, as well as to astrophysical or modeling systematics.
The analysis presented in this paper was blinded in the sense that all analysis choices were fixed and we ensured that the pipeline passed a number of predetermined validation tests before we looked at the true cosmological results. The blinding procedure and these tests are described below.

\subsection{Validation}\label{sec:validation}

In planning and executing this study, we took several steps to protect the results against possible experimenter bias, following a procedure similar to the parameter-level blinding strategy used in previous DES Y1 beyond-\lcdm analyses. Key to this were extensive simulated analyses, in which we analyzed model predictions for observables with known input parameters  as if they were data.
All analysis choices are based on these simulated analyses, including which datasets we focus on and how we report results. Before running our analysis pipeline on real data, we wrote the bulk of this paper's text, including the plan of how the analysis would proceed, and subjected that text to a preliminary stage of DES internal review.

When performing parameter estimation on the real data, we concealed the cosmology results using the following strategies:
\begin{itemize}
\item We avoided over-plotting measured data and theory predictions for observables.
\item We post-processed all chains so that the mean of the posterior distributions lay on our fiducial cosmology.
\item We do not look at model comparison measures between our split parameterization and \lcdm.
\end{itemize}
We maintained these restrictions until we confirmed that the analysis passed several sets of validation tests:
\begin{itemize}
\item We confirm that our results cannot be significantly biased by
  any one of the sample systematics adopted in our validation tests. To do this we check that the parameter estimates we report change by less than  $0.3\sigma$ when we contaminate synthetic input data with a number of different effects, including non-linear galaxy bias and a more sophisticated intrinsic alignment model. This test  is discussed \app{app:systests}.
\item We confirm that non-offset \lcdm chains give results consistent with what  Ref.~\cite{Abbott:2018wzc} reports.\footnote{The data combinations we use are slightly different than those in Ref.~\cite{Abbott:2018wzc}, so we simply require that our \lcdm results be reasonably consistent with theirs, rather than identical.}
\item We studied whether our main results are robust to changes in our analysis pipeline. We found that parameter constraints shift by less than $0.3\sigma$ when we apply more aggressive  cuts to removing non-linear angular scales, and when we use an alternative set of photometric redshifts.

  Our results did change when we replaced the intrinsic alignment model defined \eq{eq:iadef} with one where the amplitude \aia varies independently in each source redshift bin. Upon further investigation, detailed in~\app{app:realdattests},  we found that a similar posterior shift manifests in the analysis of synthetic data, so we believe that it is due to a  parameter-space projection effect rather than a property of the real DES data. We therefore proceed with the planned analysis despite  failing this robustness test, but add an examination of how intrinsic alignment properties covary with our split parameters to the discussion in \sect{sec:bigplot_results}.
\end{itemize}

After passing another stage of internal review, we then finalized the analysis by updating the plots to show non-offset posteriors, computing tension and model comparison statistics, and writing descriptions of the results. After unblinding a few changes were made to the analysis: First, we discovered that our real-data results had accidentally been run using Pantheon~\cite{Scolnic:2017caz} supernovae, so we reran all chains to include correct DES SNe data. While doing this, we additionally made a small change to our compressed Planck likelihood,  centering its Gaussian likelihood on the full Planck chain's mean parameter values, rather than on maximum-posterior sample.  This choice was motivated by the fact that sampling error in the maximum posterior estimate means that compressed likelihood is more accurate when centered on the mean. We estimate that centering on the maximum posterior sample was causing the compressed likelihood to be biased by $\sim 0.2\sigma$ relative to the mean, though we avoided looking at the direction of this bias in parameter space in order to prevent our knowledge of that direction from influencing this choice.

\subsection{Evaluating tensions and model comparison}\label{sec:postunbl}

There are two senses in which measuring tension is relevant for this analysis. First, we want to check for tension between different datasets in order to determine whether it is sensible to report their combined constraints. Second, we want to test whether our split-parameterization results are in tension with \lcdm (or \wcdm in the case of split $w$).  For both of these applications, we evaluate tension using Bayesian suspiciousness~\cite{Handley:2019wlz,Lemos:2019txn}, which we compute using {\sc anesthetic}.\footnote{\url{https://github.com/williamjameshandley/anesthetic}}~\cite{Handley:2019mfs}

Suspiciousness  $S$ is a quantity built from the Bayesian evidence ratio $R$ designed to remove dependence of the tension metric on the choice of prior. Let us define $S^{\rm dat}$ to measure the tension between two datasets $A$ and $B$.
The Bayesian evidence ratio  between  $A$ and $B$'s  constraints 
is
\lneqb
\begin{equation}\label{eq:evratio}
  R^{\rm dat} = \frac{\mathcal{Z}_{AB}}{\mathcal{Z}_{A}\mathcal{Z}_{B}},
\end{equation}\lneqe
where $\mathcal{Z}_{X}=\int d\Theta\, \mathcal{P}(\Theta|X)$  is the Bayesian evidence for dataset $X$ with posterior $\mathcal{P}(\Theta|X)$.
Generally, values of $R^{\rm dat}>1$ indicate agreement between $A$ and $B$'s constraints, while $R^{\rm dat}<1$ indicates tension, though the translation of $R$ values into tension probability depends on the choice of priors~\cite{Handley:2019wlz,Raveri:2018wln}.

The Kullback-Leibler (KL)  divergence
\lneqb
\begin{equation}\label{eq:kldiv}
  \mathcal{D}_X = \int d\Theta\,\mathcal{P}(\Theta|X)\,\log{\left[\mathcal{P}(\Theta|X)/\pi(\Theta)\right]},
\end{equation}\lneqe
measures the information gain between the prior and the posterior for constraints based on dataset $X$. The comparison between KL divergences can be used to quantify the probability, given the prior, that constraints from datasets $A$ and $B$ will agree. This information is encapsulated in the information ratio,
\lneqb
\begin{equation}\label{eq:inforatio}
\log{I^{\rm dat}} = \mathcal{D}_A + \mathcal{D}_B - \mathcal{D}_{AB},
\end{equation}\lneqe
where $\mathcal{D}_{AB}$ is the KL divergence for the combined analysis of $A$ and $B$.
To get Bayesian suspiciousness we subtract the information ratio from the Bayesian evidence:
\lneqb\begin{equation}\label{eq:datcompS}
  \log{S^{\rm dat}} = \log{R^{\rm dat}} - \log{I^{\rm dat}}.
\end{equation}\lneqe
This subtraction makes $S$ insensitive to changes in the choice of priors, as long as those changes do not significantly impact the posterior shape.
As with $R$, larger values of $S$ indicate greater agreement between datasets.

To translate this into a more quantitative measure of consistency, we use the fact that the  quantity $d-2\log{S}$ approximately follows a $\chi_d^2$ probability distribution, where $d$ is the number of parameters constrained by both datasets. In practice we determine $d$ by computing the Bayesian model dimensionality $d$~\cite{Handley:2019pqx}, which accounts for the extent to which our posterior is unconstrained (prior-bounded) in some parameter-space directions. The model dimensionality for  a single set of constraints $X$ is defined as
\lneqb
\begin{equation}\label{eq:bmd}
  \frac{\tilde{d}_X}{2} = \int d\Theta\,\mathcal{P}(\Theta|X)\,\left(\log{\left[\mathcal{P}(\Theta|X)/\pi(\Theta)\right]}\right)^2 - \mathcal{D}_X^2.
\end{equation}\lneqe
This measures the variance of the gain in information provided by $X$'s posterior. Though $\tilde{d}$ is generally non-integer, it can be interpreted as the effective number of constrained parameters.  To get the value of $d$ that we use for our tension probability calculation, we compute
\lneqb\begin{equation}\label{eq:dfortension}
  d\equiv d_{A\cap B} = \tilde{d}_A + \tilde{d}_B - \tilde{d}_{AB}.
\end{equation}\lneqe
Since any parameter constrained by either $A$ or $B$ will also be constrained by their combination, this subtraction will remove the count for any parameter constrained by only one dataset. Thus, $d$ is the effective number of parameters constrained by both datasets.
As we noted above, the quantity $d-2\log{S}$ approximately follows a $\chi_d^2$ probability distribution, so we compute the tension probability
\lneqb\begin{equation}\label{eq:tensionprob}
  p(S>S^{\rm dat}) = \int_{d-2\log{S}}^{\infty}\chi_d^2(x)\,dx,
\end{equation}\lneqe
which quantifies the probability that the datasets $A$ and $B$ would be more discordant than measured.
If in our analysis we find $p(S>S^{\rm dat})<5\%$, we will consider the two datasets to be in tension and will not report parameter constraints from their combination.

We will also use the Bayesian Suspiciousness in order to perform model comparison. One can interpret the Bayesian Evidence Ratio and Suspiciousness defined in \eqs{eq:evratio}-~(\ref{eq:datcompS}) as a test of the hypothesis that datasets $A$ and $B$ are described by a common set of cosmological parameters as opposed to two independent sets. That can be directly translated into what we would like to determine: are the data in tension with a single set of parameters describing both growth and geometric observables?
We therefore compute
\lneqb\begin{align}
  R^{\rm mod} &= \mathcal{Z}_{\rm \lcdm}/\mathcal{Z}_{\rm mod}\\
  I^{\rm mod} &= \mathcal{D}_{\rm \lcdm} - \mathcal{D}_{\rm mod} \\
  \log{S^{\rm mod}} &= \log{R^{\rm mod}} - \log{I^{\rm mod}} \label{eq:modcompS}
\end{align}\lneqe
We use the label ``mod'' to identify these as model comparison statistics.  As before we translate this into a tension probability by computing the Bayesian model dimensionality,
\lneqb\begin{equation}
  d = d_{\rm mod} - d_{\lcdm},
\end{equation}\lneqe
and integrating the expected $\chisq_d$ distribution as in \eq{eq:tensionprob}. The resulting  quantity $p(S>S^{\rm mod})$ measures the probability to exceed the observed  tension between growth and geometric observables.

To convert a probability $p$ to an equivalent $N\sigma$ scale, we compute $N$ such that $p$ is the probability that $|x|>N$ for a standard normal distribution,
\lneqb
\begin{align}
  p &= 1 - 2\int_0^N (2\pi)^{-1/2}e^{-x^2/2}\,dx
  = \erfc{\left(\tfrac{N}{\sqrt{2}}\right)},\\
  N&= \sqrt{2}\erfc^{-1}(p).\label{eq:doubletailsigma}
  \end{align}
\lneqe
Unless otherwise noted, this double-tail equivalent probability is what will
be used to convert probabilities to $N\sigma$. In the specific case
  when we are testing in \sect{sec:consistency} whether the difference between the corresponding
  growth  and geometry parameter is greater than zero, a single-tail
  probability is relevant instead; in that case, we simply multiply $p$ in \eq{eq:doubletailsigma} by a factor of two.

\section{Results: Split parameters}\label{sec:results}

Here we present our main results, which are constraints on split parameters and an assessment of  whether or not the data are consistent with $\paramsgrow=\paramsgeo$. \sect{sec:results_som} reports results for splitting \om (with $w=-1$), while results for splitting both \om and $w$ are presented in \sect{sec:results_sw}. We summarize the results in \sect{sec:resultssummary}, reporting constraints, tension metrics, and model comparison statistics in \tab{tab:results}.

All datasets considered fulfill the $p(S>S^{\rm dat})\geq 0.05$ prerequisite set in \sect{sec:postunbl} for reporting combined constraints. Note, however, that while this is strictly true, the \lcdm constraints from DES and Ext-geo, as well the split \om constraints from DES and Ext-all  are found to have tensions at  the $2\sigma$ threshold. Thus, while we will report these combined results, they should be interpreted with caution.

Note that while one might assume that the $2\sigma$ tension found between DES and Ext-geo constraints in \lcdm 
is related to the familiar Planck-DES  \sigeight offset, this is not necessarily the case. This is  because the \sigeight tension is generally studied in terms of the constraints from the full CMB power spectrum, while we are only  using limited, geometric information from the CMB. When we do examine  marginalized \lcdm posteriors (not shown), we find substantial overlap between the $1\sigma$ regions of the marginalized DES and Ext-geo constraints on \sigeight. 
Similarly, we find no obvious incompatibility between DES and Ext-geo constraints on any other individual parameter.  
This $2\sigma$ tension therefore appears to be related to the higher-dimensional properties of the two posteriors.

\subsection{Splitting \om}\label{sec:results_som}

\begin{figure}
  \centering
  \includegraphics[width=\linewidth]{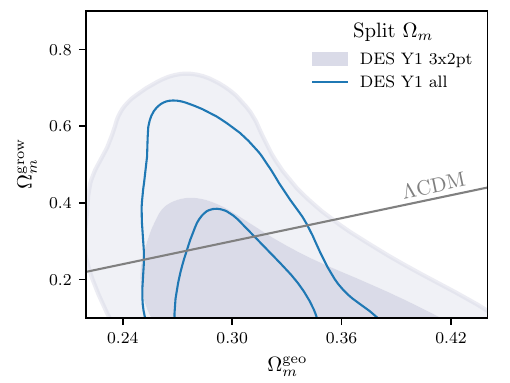}
  \includegraphics[width=\linewidth]{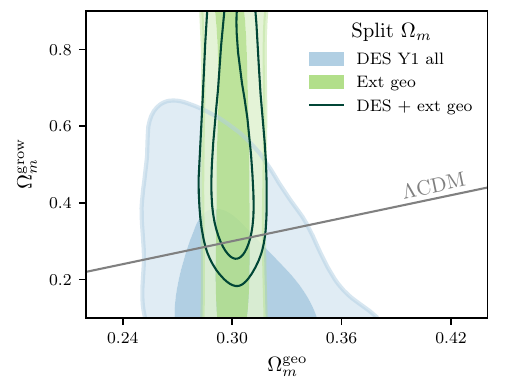}
  \includegraphics[width=\linewidth]{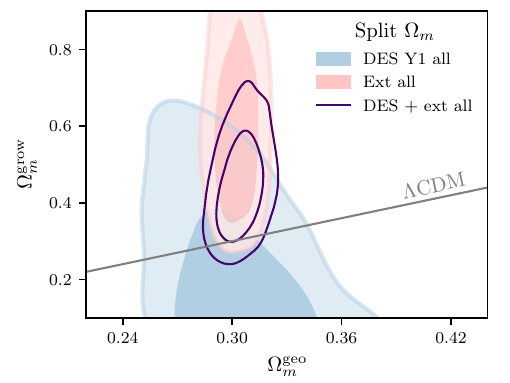}
\caption{The 68 and 95\% confidence regions for \omgrow and \omgeo for our  various data combinations. The diagonal gray lines show where $\omgrow=\omgeo$. Note that the three plots have the same axis ranges, and that vertical axes cover a much larger range of values than the horizontal axes. The blue outline-only contours in the top plot are the same as the shaded blue contours in the other plots. }
\label{fig:sOm_results}
\end{figure}

\fig{fig:sOm_results} shows the 68 and 95\% confidence regions for \omgrow and \omgeo for various data combinations.  We study three different comparisons: a comparison between our fiducial DES dataset and a version without the BAO and SNe in the top panel; DES plus external geometric (DES+Ext-geo) data in the middle panel, and DES plus external data including  RSD (Ext-all) in the bottom panel.  The diagonal gray line corresponds to $\omgrow=\omgeo$. Marginalized parameter constraints and tension metrics for both data combination and model comparison are reported in \tab{tab:results}.

Looking at DES-only results in the top panel, we find that, as expected including the (geometric) DES BAO and SNe likelihoods tightens the constraints on \omgeo but only weakly affects \omgrow.
We find that constraints on \omgeo are much stronger than those on \omgrow for both the \mpp-only and the fiducial DES constraints.  In fact, the DES constraints on \omgeo are only slightly weaker than \lcdm constraints on \om, implying that  most of DES' constraining power is derived from geometric information. This might be surprising, since one might expect a LSS survey to have more growth sensitivity.
However, it is consistent with the findings summarized in Ref.~\cite{Zhan:2008jh}, which discusses how distance and growth factor measurements can place comparable constraints on the dark energy equation of state when other cosmological parameters are held fixed~\cite{Simpson:2004rz,Zhang:2003ii}, but the growth weakens when one marginalizes over more parameters~\cite{Abazajian:2002ck,Knox:2005rg}.  The fact that the confidence regions intersect the $\omgrow=\omgeo$  line but are asymmetrically distributed around it is reflected in the Bayesian Suspiciousness measurement of $1.5\sigma$ tension with \lcdm.

In the middle panel of \fig{fig:sOm_results} we show the combination of the
DES data with external  geometric measurements from the CMB and BAO (Ext-geo).
As
expected, the external geometric data alone  put tight constraints on \omgeo
but do not constrain \omgrow at all. The combined constraints on \omgeo are straightforwardly dominated by
those from the external data, while the DES+Ext-geo constraints on \omgrow are counterintuitively bounded from below but not above.
To understand the appearance of the lower bound, note that the  DES-only measurement of a given late-time density fluctuation amplitude  allows arbitrarily small values of \omgrow because little or no structure growth over time can be compensated by a large primordial amplitude \as.
Adding the Planck constraints provides an early-time anchor for \as, and therefore requires \omgrow to be above some minimal value in order to account for the evolution of structure growth between recombination and the redshifts probed by DES. The reason DES'  upper bound on \omgrow does not translate to the DES+Ext-geo constraints can also be understood in terms of degeneracies in our model's larger parameter space. We will explore this in more detail in  \sect{sec:bigplot_results}.

Finally, the bottom panel of \fig{fig:sOm_results} shows constraints from DES and Ext-all, which adds BOSS RSD constraints on growth to the previously considered external geometric measurements. We see that compared to the middle panel's Ext-geo results, adding RSD allows Ext-all to place a lower bound on \omgrow, and when combined with DES, \omgrow is bounded on both sides.  The fact that there is not very much overlap between the DES and Ext-all contours, with Ext-all preferring somewhat higher \omgrow than DES, reflects their weak $ 2\sigma$ tension.
The shape of the Ext-all constraints here, as well as how DES adds information, is related to a degeneracy between \omgrow and \mnu, which we will discuss further in \sect{sec:bigplot_results}.

\subsection{Splitting \om and $w$}\label{sec:results_sw}

\begin{figure}
  \centering
  \includegraphics[width=\linewidth]{{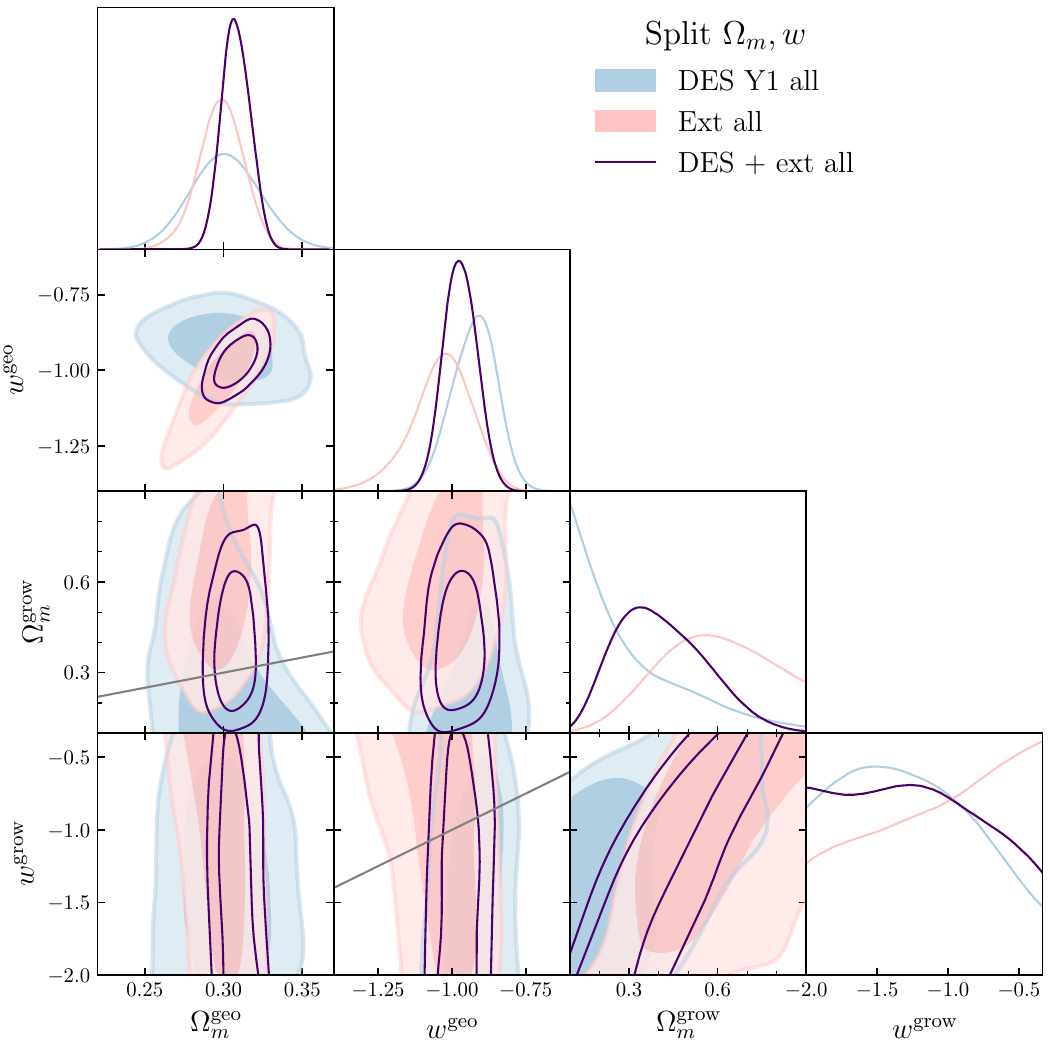}}
\caption{Marginalized constraints from DES and external data when both \om and $w$ are split. The diagonal panels show normalized one-dimensional marginalized posteriors, while the off-diagonal panels  show 68\% and 95\% confidence regions. Solid gray lines show the \wcdm parameter subspace where $\omgrow=\omgeo$ and $\wgrow=\wgeo$.}
\label{fig:sOm-sw_results-bigplot}
\end{figure}

\fig{fig:sOm-sw_results-bigplot} shows the 68\% and 95\% confidence contours when splitting both \om and $w$ for DES+Ext-all constraints, showing the parameters \omgeo, \omgrow, \wgeo, and  \wgrow. The most notable feature is the strong degeneracy between the two growth parameters, \omgrow and \wgrow. We interpret this to mean that while DES+Ext-all can separately constrain growth and geometry, the data cannot distinguish  between \om-like and $w$-like deviations from the structure growth history expected from \wcdm. This behavior is also consistent with   Ref.~\cite{Linder:2005in}'s finding that, for a given $\om(z)$, $w$ only weakly effects growth rates. This makes it unsuprising that it is difficult to robustly constrain \wgrow separately from \omgrow.

Because of this degeneracy,  even using our most informative ``DES+Ext-all'' data combination \wgrow is unconstrained, and the upper and lower bounds placed on \omgrow are entirely dependent on the choice of prior for \wgrow.
As discussed in  \app{sec:syndat}, our analyses of simulated data showed that projection effects associated with this degeneracy significantly affect the  one-dimensional marginalized constraints on both \omgrow and \wgrow. Because of this we do not report parameter constraints for this model.

\subsection{Consistency with $\paramsgrow=\paramsgeo$}\label{sec:consistency}

Ultimately the question we would like to ask is whether the results above are consistent with \lcdm, or with \wcdm, in the case where we split both \om and $w$. There are several ways we can assess this. We  begin simply by looking at the two-dimensional confidence regions shown in  \figs{fig:sOm_results} and~\ref{fig:sOm-sw_results-bigplot}, noting whether or not they intersect the lines corresponding to \lcdm (in  \fig{fig:sOm_results}) and \wcdm (in  \fig{fig:sOm-sw_results-bigplot}).  We see that when we split \om, the  68\% confidence intervals for  DES and DES+Ext-geo intersect the  $\omgrow=\omgeo$ line, while that of DES+Ext-all just touches the \lcdm line, preferring $\omgrow>\omgeo$. When we split both \om and $w$, both the $\omgrow=\omgeo$ and $\wgrow=\wgeo$ lines goes directly through the DES+Ext-all 68\% confidence intervals.

To assess consistency with $\paramsgrow=\paramsgeo$ in our full parameter space, we use Bayesian Suspiciousness $S^{\rm mod}$, as described in \eq{eq:modcompS} of \sect{sec:postunbl}. As we did when we used Suspiciousness to evaluate concordance between datasets, we use  $p(S>S^{\rm mod})$ to report the probability to exceed the observed Suspiciousness, and ``1-tail equiv. $\sigma$'' as the number of normal distribution standard deviations with equivalent probability. Here, larger $S^{\rm mod}$, smaller $p(S>S^{\rm mod})$, and larger $\sigma$ indicate more  tension with $\paramsgrow=\paramsgeo$. Numbers for all of these quantities are shown in \tab{tab:results}. According to this metric, when we split \om we find the DES-only results to have a  $1.5\sigma$ tension with \lcdm. This becomes $1.9\sigma$ for DES+Ext-geo, and $1.0\sigma$ for DES+Ext-all. When we split both \om and $w$ we find tensions with \wcdm to be $1.6\sigma$ for DES-only and $1.4\sigma$ for DES+Ext-all.

\begin{figure}
  \centering
  \includegraphics[width=\linewidth]{{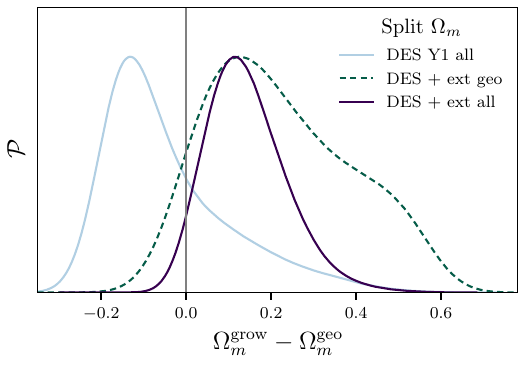}}
\caption{Marginalized posterior of the difference $\omgrow-\omgeo$, from fitting the split-\om model to the DES, DES+Ext-geo, and DES+Ext-all data combinations.}
\label{fig:diff-sOm}
\end{figure}

As another way of quantifying compatibility of the split-\om constraints with \lcdm,  in \fig{fig:diff-sOm} we show the marginalized posterior for the difference $\omgrow - \omgeo$.   When we assess the fraction of the posterior volume above and below 0, we find that the fraction of the posterior volume with $\omgrow>\omgeo$ is 30\% for DES-only, equivalent to a normal distribution single-tail probability of $0.5\sigma$. These numbers become 91\% ($1.3\sigma$) for DES+Ext-geo, and 95\% ($1.6\sigma$) for  DES+Ext-all.

We note two points of caution in interpreting the $\omgrow-\omgeo$ marginalized posterior. First, because of the difference in constraining power on \omgrow and \omgeo there is some asymmetry expected in these marginalized posteriors even if the data are consistent with \lcdm. This can be seen in \fig{fig:systests_differences} of \app{app:systests}, which shows versions of this plot for synthetic data generated with $\omgrow=\omgeo$. Additionally, the  posterior distribution is impacted by the priors on \omgrow and \omgeo. While the {\sc GetDist} software allows us to correct for the impact of hard prior boundaries for the parameters we sample over, it is unable to do so for derived parameters. This means that in cases where the shape of the posterior is influenced by the prior boundary of e.g. \omgrow, this  will necessarily affect the shape of the marginalized posterior for $\omgrow-\omgeo$. 
Accounting for these caveats and comparing to the simulated results in \app{app:systests}, we see that the DES+Ext-all probability distribution is shifted to  higher $\omgrow-\omgeo$ than  was found in simulated analyses.
The DES-only and DES+Ext-geo distributions do not appear to be significantly different from what might be expected given parameter space projection effects in \lcdm.

\subsection{Summary of main results}\label{sec:resultssummary}

\begin{table}
  \begin{center}
    \caption{Summary of results.  Parameter errors quoted are 68\% confidence intervals, and  $\tilde{d}$ is the Bayesian model dimensionality. The quantity $S$ is Bayesian Suspiciousness, with the superscript ``dat'' denoting an assessment of tension between two datasets, and ``mod'' denoting model comparison.
    }
    \label{tab:results}
    \renewcommand{\arraystretch}{1.3} 
    \begin{tabular}{c|ccc}

      \textbf{\lcdm} & DES &  DES+Ext-geo & DES+Ext-all  \\ \hline
      \om  &  $0.296^{+0.020}_{-0.022}$ & $0.301^{+0.009}_{-0.008}$ & $0.302^{+0.007}_{-0.008}$\\
      $\tilde{d}$  & $ 14.0\pm 0.7 $ & $ 18.2\pm 0.8$ & $15.8 \pm 0.8$ \\\hline

      $\log S^{\rm dat}$ & - &  $ -1.8\pm 0.3$ & $ -0.8\pm0.2 $\\
      $p(S>S^{\rm dat})$ & - & $ 0.04\pm  0.03 $ & $ 0.25 \pm 0.05 $  \\
      equiv. $\sigma$ & - & $  2.0\pm 0.4$ & $1.2\pm 0.1$ \\\hline
      \multicolumn{4}{c}{}\\

      \textbf{Split \om} & DES & DES+Ext-geo & DES+Ext-all  \\\hline
      $\omgrow - \omgeo$ &  -  &  $0.126^{+0.228}_{-0.129}$ & $0.116^{+0.100}_{-0.084}$  \\
      \omgeo & - 
      &  $0.300^{+0.009}_{-0.008}$ &  $0.304^{+0.009}_{-0.008}$\\
      \omgrow &  -  &  $0.425^{+0.232}_{-0.131}$ &  $0.421^{+0.102}_{-0.089}$\\
      $\tilde{d}$  & $ 15.0 \pm 0.7$ & $18.1 \pm 0.9$ & $ 19.8\pm 1.0 $  \\\hline

      $\log S^{\rm dat}$ & - & $-1.1 \pm -0.2 $  & $ -2.0\pm 0.3$\\
      $p(S>S^{\rm dat})$ & - & $0.13\pm 0.06$& $ 0.05 \pm 0.03$ \\
      equiv. $\sigma$ & - & $1.5\pm 0.3$  & $ 2.0 \pm 0.3 $  \\\hline

      $\log S^{\rm mod}$ & $ -0.6\pm 0.2 $ & $ -1.3\pm 0.2 $ & $-0.4 \pm 0.3$ \\
      $p(S>S^{\rm mod})$ & $  0.14 \pm 0.08 $ & $ 0.06 \pm 0.05 $ & $0.31 \pm 0.07 $ \\
      equiv. $\sigma$ & $ 1.5 \pm 0.4 $ & $ 1.9\pm 0.4 $ & $ 1.0\pm  0.1$ \\\hline

      $p{(\omgrow>\omgeo)}$& 0.30 & 0.91 & 0.95\\
      1-tail equiv. $\sigma$ & 0.5 & 1.3 & 1.6\\\hline

      \multicolumn{4}{c}{}\\

      \textbf{\wcdm} & DES &  DES+Ext-all  \\\cline{1-3}
      \om & $0.292^{+0.022}_{-0.022}$ & $0.301^{+0.009}_{-0.008}$ \\
      $w$  & $-0.911^{+0.073}_{-0.076}$ & $-0.997^{+0.048}_{-0.050}$ \\
      $\tilde{d}$  & $16.3 \pm 0.8 $ &  $17.6 \pm 0.8$\\\cline{1-3}

      $\log S^{\rm dat}$ & - & $-1.5 \pm 0.2  $\\
      $p(S>S^{\rm dat})$ & - &  $ 0.17\pm 0.04  $ \\
      equiv. $\sigma$ & - &  $ 1.4\pm 0.1$ \\\cline{1-3}
      \multicolumn{4}{c}{}\\

      \textbf{Split \om, $w$} & DES & DES+Ext-all  \\\cline{1-3}
      $\tilde{d}$  &  $ 15.2 \pm 0.7 $ & $ 18.2\pm 0.9$ \\\cline{1-3}

      $\log S^{\rm dat}$ & - &   $-2.2 \pm 0.2 $\\
      $p(S>S^{\rm dat})$ & - &  $ 0.09\pm 0.03  $ \\
      equiv. $\sigma$ & - & $ 1.7\pm 0.2$ \\\cline{1-3}

      $\log S^{\rm mod}$ & $-0.5 \pm 0.2$  & $ -0.6 \pm 0.2$ \\
      $p(S>S^{\rm mod})$ & $0.11 \pm 0.07$  &  $0.15 \pm 0.09$\\
      equiv $\sigma$ & $ 1.6\pm 0.4$  &  $ 1.4\pm 0.4$\\\cline{1-3}
    \end{tabular}
  \end{center}
\end{table}

The results discussed in this section are summarized in  \tab{tab:results}. In it, for the split \om model
 we show one-dimensional marginalized constraints on \omgrow and \omgeo from DES+Ext-geo and DES+Ext-all, along with  \lcdm and \wcdm  constraints for comparison. For each parameter we show two-sided errors corresponding to the 68\% confidence interval one-dimensional marginalized posterior.
Because we expect the one-dimensional marginalized posteriors  to be subject to significant projection effects for  DES-only constraints on the split \om model and for the DES+Ext-all constraints when splitting both \om and $w$, as discussed in \sect{sec:analysis} and \app{sec:syndat} we do not report parameter bounds for those cases.

For all model-data combinations considered we use Bayesian Suspiciousness  as  defined in \sect{sec:postunbl} to report data tension and model comparison statistics. In \tab{tab:results}, $\tilde{d}$ is the Bayesian model dimensionality (\eq{eq:bmd}) quantifying  the effective number of parameters constrained, $S^{\rm dat}$ is the Bayesian suspiciousness assessing agreement between pairs of datasets (\eq{eq:datcompS}), and  $S^{\rm mod}$ is the model-comparison Bayesian Suspiciousness (\eq{eq:modcompS}), quantifying tension or agreement with $\paramsgrow=\paramsgeo$. The quantities $p(S>S^X)$, for $X\in[{\rm dat},{\rm mod}]$ is the probability that a random realization exceeds the observed Suspiciousness $S^X$, and ``equiv. $\sigma$'' translates that probability into the number standard deviations with an equivalent double-tail probability for a normal distribution (\eq{eq:doubletailsigma}). Large $S$, small $p$, and large equivalent $\sigma$ indicate tension, while small $S$, large $p$, and small equivalent $\sigma$ indicate concordance. For all quantities the numbers quoted in \tab{tab:results} are the mean and standard deviation from sampling error reported by {\sc Anesthetic}.

As an alternative model-comparison statistic for the split-\om model, we additionally report $p{(\omgrow>\omgeo)}$, the fraction of the posterior volume with $\omgrow>\omgeo$. For this part of the table, the ``equiv. $\sigma$'' is the number of normal distribution standard deviations with equivalent single-tail probability.

\section{Results: Impact of growth-geometry split on other parameters}\label{sec:bigplot_results}

\begin{figure*}
  \centering
  \includegraphics[width=.8\linewidth]{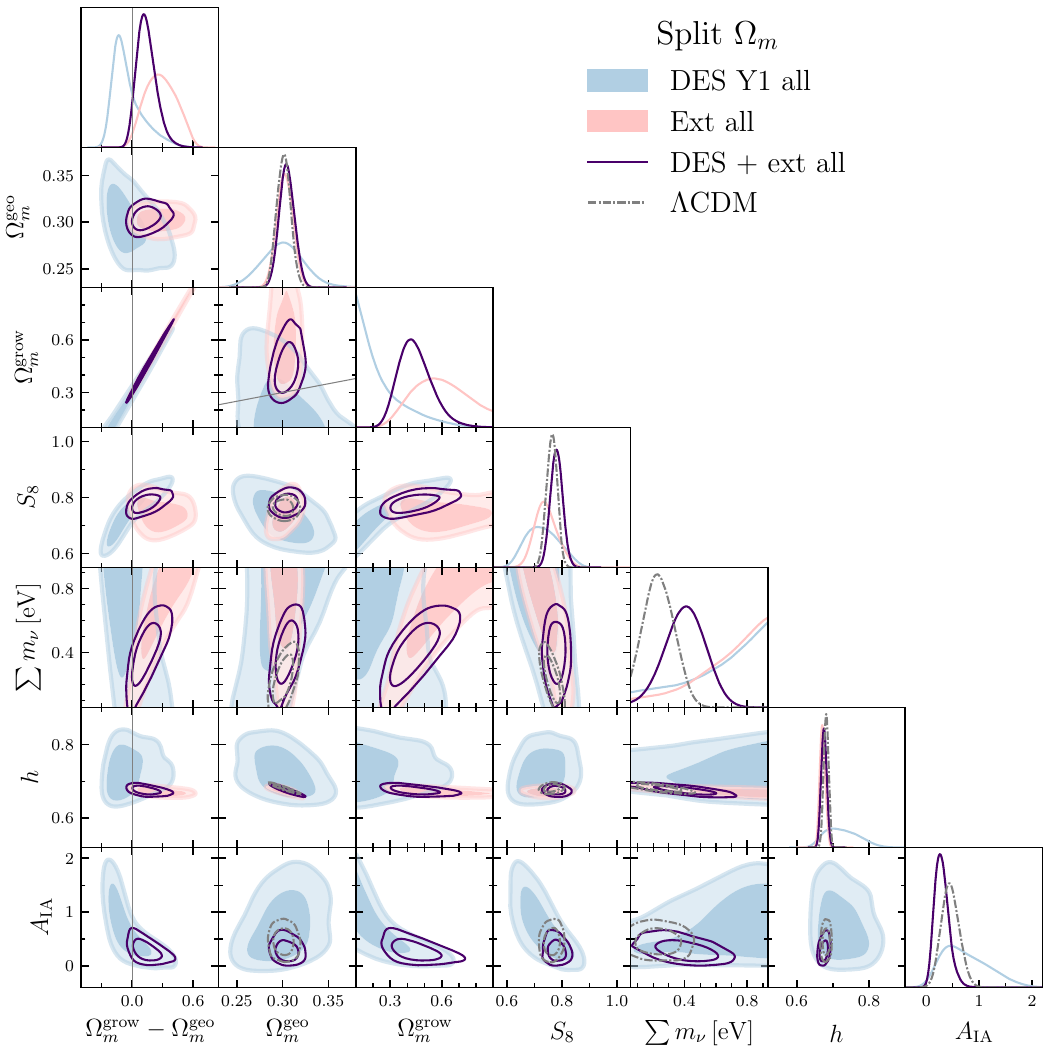}
\caption{Constraints from the DES and the Ext-all external dataset, which includes the compressed Planck likelihood, BOSS DR12 BAO, and BOSS DR12 RSD. The off-diagonal panels show the 68 and 95\% confidence intervals for each data combination, while the diagonal panels show normalized one-dimensional marginalized  posteriors on parameters. DES-only results are shown in blue, Ext-all results are pink, and their combination is shown using unshaded purple contours. The gray dashed curves show  DES+Ext-all constraints in \lcdm and the gray solid lines show where $\omgrow=\omgeo$. }
\label{fig:degeneracies_sOm_p5br}
\end{figure*}

Here we explore how our split parameterization, focusing on  splitting only \om, affects the inference of other cosmological parameters. In this discussion we will primarily reference \fig{fig:degeneracies_sOm_p5br}, which shows two-dimensional marginalized posteriors of  DES+Ext-all constraints on \omgrow, \omgeo, the difference $\omgrow-\omgeo$, \mnu, $S_8\equiv\sigeight\sqrt{\omgeo/0.3}$, $h$, and \aia. For comparison, we also show a DES+Ext-geo version of this plot in  \fig{fig:degeneracies_sOm_p5b} of \app{sec:moreplots}. 
We use this higher dimensional visalization of the posterior to characterize how  additional degrees of freedom  in the relationship between expansion history and structure growth  change considerations in cosmological analyses, both in terms of how we model of  astrophysical effects ($\mnu$, \aia) and in terms of commonly studied tensions (\seight, $h$).

In the off-diagonal panels of \fig{fig:degeneracies_sOm_p5br}, 68\% and 95\% confidence regions are shown for DES-only as blue shaded contours, Ext-all as pink shaded contours, and the combination DES+Ext-all as dark purple outlines. The diagonal panels show normalized  one-dimensional marginalized posteriors for each parameter. Solid gray lines show the \lcdm subspace where \omgrow=\omgeo, and grey dashed lines show the DES+Ext-all posterior for \lcdm.

\subsection{Effect of split on neutrino mass}\label{sec:mnu_results}

Because the combination of Planck, BOSS BAO, and BOSS RSD are able to tightly constrain cosmological parameters in \lcdm, it may be surprising that DES adds information at all when combined with the Ext-all data. Looking at \fig{fig:degeneracies_sOm_p5br}, we see that it does so because the external data  exhibits a significant degeneracy between  \omgrow and the sum of neutrino masses \mnu. The Ext-all degeneracy occurs because changes in \mnu and \omgrow  have competing effects on the matter power spectrum: higher neutrino mass suppresses structure
formation at small scales ($k\gtrsim 10^{-2}h\text{ Mpc}^{-1}$), while raising \omgrow results produces more late-time structure. DES data adds constraining power because it provides an upper bound on \omgrow which breaks that degeneracy. 

Looking at the marginalized constraints on \mnu, we see that both the DES+Ext-all (\fig{fig:degeneracies_sOm_p5br}) and DES+Ext-geo (\fig{fig:degeneracies_sOm_p5b}) constraints produce a detection of neutrino mass at $\mnu=0.4\pm 0.1\,\text{eV}$, which is significantly higher than the upper bounds  obtained from the combined analysis of BOSS DR12  and the full Planck temperature and polarization power spectra~\cite{Alam:2016hwk,Aghanim:2018eyx}. The DES-only posterior gives a weak lower bound on neutrino mass, though we suspect that this may be at least in part caused by parameter-space projection effects. 
In \lcdm, the Ext-all constraints on \mnu become an upper bound of $\mnu<0.45\text{ eV}$ at 95\% confidence, which is is consistent with the BOSS results (though weaker because we do not use the full Planck likelihood), while the DES preference for high \mnu remains. This causes the DES+Ext-all \lcdm posterior, shown as a gray dashed line in \fig{fig:degeneracies_sOm_p5br},  to peak at  $\mnu= 0.2\pm 0.1\,\text{eV}$. 

To begin interpreting the preference for high \mnu, we can look at the \mnu-\omgrow panel of \fig{fig:degeneracies_sOm_p5br} and  note that 
the Ext-all constraints exhibit a preference  for the high-\mnu, high-\omgrow part of parameter space. That preference combined with the DES upper bound on \omgrow likely drives the $2\sigma$ tension between Ext-all and DES, and it appears to be responsible for pulling the combined DES+Ext-all constraints away from the \lcdm $\omgrow=\omgeo$ line.

\begin{figure}
  \centering
  \includegraphics[width=\linewidth]{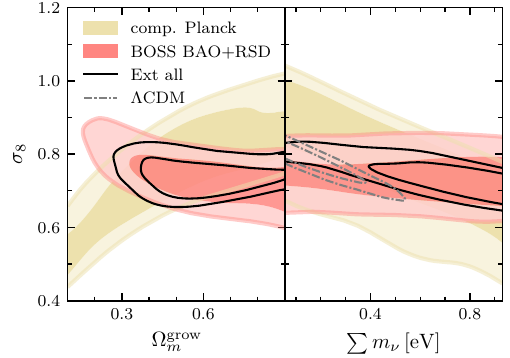}
\caption{Constraints on parameters most relevant for describing late-time growth, shown for the datasets that make up Ext-all. Contours show the 68\% and 95\% confidence regions for the compressed Planck likelihood in yellow and for BOSS DR12 BAO and RSD in orange. The unshaded black contours correspond to Ext-all, and are the same as the pink contours in other Figures.  Gray dashed contours show \lcdm results for Ext-all. }
\label{fig:growthrectangle_extal}
\end{figure}

It is instructive to examine how the constituent Planck and BOSS likelihoods combine to produce the Ext-all contours.  We show this in \fig{fig:growthrectangle_extal}, with the compressed Planck posterior  in yellow, BOSS BAO+RSD in orange, and their combination, Ext-all, as black outlines.  The compressed Planck likelihood approximately defines a plane in the \sigeight-\mnu-\omgrow parameter space because Planck's measurement of \as can be extrapolated forward to predict \sigeight, but the effects of \omgrow and \mnu on late-time structure growth loosen that predictive relationship. The BOSS data probe late-time structure more directly, so the combined BAO and RSD results can be thought of as roughly providing a measurement of \sigeight that is insensitive to \mnu and only weakly dependent on \omgrow.

Putting all of this together, we see that the shape of the Ext-all posterior strongly depends on the relationship between Planck's measurement of \as,  BOSS's measurement of \sigeight, as well as the extent to which late-time degrees of freedom impact how deterministically Planck's \as constraint maps to \sigeight. For example: if the Planck \as constraints were lowered slightly, or the BOSS \sigeight constraints were raised, this would move the Ext-all constraints  towards lower \mnu and consequently, lower \omgrow. The DES+Ext-geo constraints have a similar property: we can see in the \seight-\mnu panel of \fig{fig:degeneracies_sOm_p5b} that slight relative changes to the Planck \as or DES \seight constraints can have a significant impact on the  \mnu posterior.  In other words, our results' preference for high \mnu (and consequently, high \omgrow) can be interpreted as a manifestation of the early-versus-late-Universe \sigeight tension discussed in the Introduction.

Our findings here are in line with several previous studies which report a preference for $\mnu\sim 0.3\,\text{eV}$ when modeling degrees of freedom affecting structure growth are introduced to combined CMB and LSS analyses. These include the  growth-geometry split analysis of Refs.~\cite{Ruiz:2014hma,Bernal:2015zom}, as well as examinations of  neutrino mass in conjunction with $A_{\rm Lens}$~\cite{Beutler:2014yhv,McCarthy:2017csu} (which describes the amount of lensing-induced smoothing of  the CMB power spectrum), time-dependent dark energy~\cite{Poulin:2018zxs}, and modified gravity~\cite{Dirian:2017pwp}.
Notably, however, these results are in contrast  with those documented in Fig.~19 of the official BOSS DR12 analysis paper~\cite{Alam:2016hwk}, which show that BOSS DR12 BAO and RSD combined with Planck temperature and polarization are able to constrain $\mnu<0.25\,\text{eV}$ at 95\% confidence,  even when marginalizing over $A_{\rm Lens}$ and a free amplitude multiplying \fsig.
Our Ext-all constraints are weaker than this because using a compressed Planck likelihood  causes  us to lose information about a degeneracy between \mnu and the shift parameter $R$ that is present in the full likelihood (which in the BOSS analysis is broken by BAO angular diameter distance measurements), and potentially also because  our choice of priors requires $\mnu>0.06\text{ eV}$ while BOSS uses $\mnu>0 \text{ eV}$.

\begin{figure}
  \centering
  \includegraphics[width=\linewidth]{{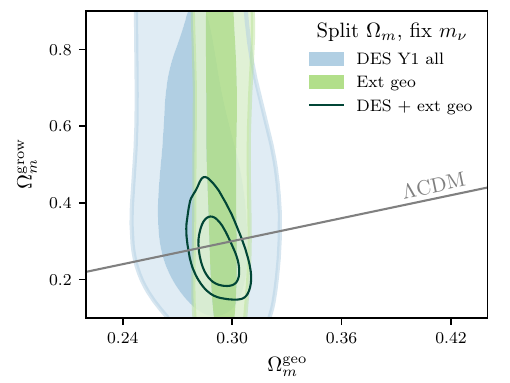}}
  \includegraphics[width=\linewidth]{{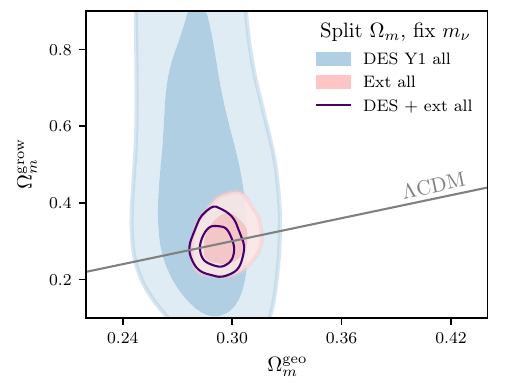}}
\caption{The same combined DES+Ext-geo (top) and DES+Ext-all (bottom) constraints as the second and third panels of \fig{fig:sOm_results}, but with the sum of neutrino masses fixed to $0.06$ eV.}
\label{fig:fixnu-da-sp5br-small}
\end{figure}

To explore how our results would be affected by 
tighter \mnu constraints, in  \fig{fig:fixnu-da-sp5br-small} we show  DES+Ext-geo and DES+Ext-all constraints on \omgrow and \omgeo when the sum of neutrino masses is fixed to its minimal value, 0.06~eV. Additionally, in \app{app:fixmnu} \fig{fig:numodcomp_extall} shows how either fixing \mnu or requiring $\omgrow=\omgeo$ alters the Ext-all constraints (without DES data), and  \tab{tab:results_fixmnu} reports fixed-\mnu versions of our data and model tension metrics.   We find that assuming minimal neutrino mass allows us to constrain \omgrow with either  DES+Ext-geo alone or just the Ext-all data, and that the fixed-neutrino-mass DES+Ext-all constraints are dominated by information from the external data. For all data combinations, fixing neutrino mass improves the agreement between datasets, and the split-\om constraints become  consistent with \lcdm  at the $<1\sigma$ level.

\subsection{Effect of split on \seight}\label{sec:sigeight_results}
In  examining the effect of the growth-geometry split parameterization on \sigeight,
we can orient ourselves by making a few observations. First, as noted in \sect{sec:modeling-mpp}, the usual
negative degeneracy between \om and \sigeight seen in weak-lensing analyses
appears in the DES-only constraints here as a degeneracy between  \omgeo (which appears in the lensing prefactor of the lensing kernel) and \sigeight. Thus, to more easily compare to results in other papers, in \fig{fig:degeneracies_sOm_p5br} we show constraints on $S_8=\sigeight\sqrt{\omgeo/0.3}$.

In contrast, the DES-only constraints on \omgrow and \sigeight are positively
correlated. This might seem counterintuitive because changing \omgrow and changing \as have similar effects on the matter power spectrum, and we are used to thinking of
\sigeight as  equivalent to \as. However, it is important to remember that splitting growth and geometry breaks our usual intuition about the one-to-one relationship between \as and \sigeight.  
While \omgrow and \as do
indeed have a negative degeneracy (see
\fig{fig:explaining-projection-effects}), \omgrow and \sigeight do not. Because \sigeight is a derived parameter obtained by integrating the power
spectrum, and increasing \omgrow raises the amplitude of the power spectrum, if all other parameters are fixed, raising \omgrow will produce an increase in \sigeight.  Thus, the degeneracy we find between \omgrow and \sigeight is expected for the same reason that we generally expect a positive correlation between  \as and \sigeight.

Compared to \lcdm,  splitting \om has relatively little effect on the DES+Ext-all \seight constraints, while it weakens and shifts those from DES and DES+Ext-geo. These \seight values remain consistent with the DES \yone \lcdm result of $0.773^{+0.026}_{-0.020}$ and below the  full Planck \lcdm constraint of $0.842^{+0.027}_{-0.025}$~\cite{Abbott:2017wau}.
We note that the DES+Ext-geo constraint on \seight  moves upwards enough to be consistent with the \lcdm Planck result, while the DES estimate moves further away --- albeit, more by reducing the lower bound than by ruling out values preferred in by DES \lcdm. This behavior is qualitatively similar to the \seight constraint in the  DES \yone for the $\Sigma-\mu$ modified gravity model analysis of Ref.~\cite{Abbott:2018xao}.

We find no disagreement between DES and the external data's marginalized \seight constraints, which might suggest that our growth-geometry split is able to resolve the \sigeight tension.
However, this interpretation is confounded by the fact that we use more limited information from Planck than standard analyses. The compressed Planck likelihood we use does not contain any information from the lensing smoothing of CMB power spectrum peaks, it is unable to constrain neutrino mass on its own, and so it exhibits a negative \seight-\mnu degeneracy. This causes the DES and Ext-geo  marginalized posteriors on \seight to be compatible even in \lcdm.
In other words, we can concretely say that the Planck preference for high \seight compared to DES and other probes of LSS relies on information from the CMB that is not included in the geometry-only compressed likelihood.


\subsection{Effect of split on $H_0$}\label{sec:h0_results}

We find that in our split parameterization, the constraints on $h\equiv H_0/(100\, \text{km}\,\text{s}^{-1}\,\text{Mpc}^{-1})$  do not significantly change relative to what they are in \lcdm.
The Planck likelihood provides nearly all the information on $h$, with its $\om h^2$ constraint manifesting as a tight ellipse in the \omgeo-$h$ planes of both \fig{fig:degeneracies_sOm_p5br} and~\ref{fig:degeneracies_sOm_p5b}. This suggests that non-standard structure growth will have little impact on the value of the Hubble constant inferred from the data we consider, and therefore is a poor candidate for resolving the $H_0$ tension.

\subsection{Effect of split on \aia}\label{sec:ia_results}

Finally, we examine how opening up our split parameter space impacts constraints on the amplitude of intrinsic alignments. We can see in \fig{fig:degeneracies_sOm_p5br} that there is a significant negative degeneracy between \omgrow and \aia present in the DES posterior. This occurs because the factor of $\rho_{m0}$ in \eq{eq:iadef} makes the contribution of intrinsic alignments signal proportional to the product $\omgrow \aia$.  As we discuss in  \sect{sec:validation} and \app{sec:syndat}, we believe this degeneracy is why our growth-focused beyond-\lcdm parameterization is more sensitive  to assumptions about the redshift dependence of intrinsic alignments than the other extensions to \lcdm considered in  Ref.~\cite{Abbott:2018xao}.

Like the previous DES Y1 papers~\cite{Abbott:2017wau,Samuroff:2018xuo,Abbott:2018ydy}, we are not able to constrain the  redshift power-law slope $\alpha_{IA}$, but we are able to constrain the amplitude \aia. For comparison, in \lcdm, our DES, DES+Ext-geo, and DES+Ext-all posteriors all give  $\aia=0.4\pm0.2$, which is compatible with (and about twice as constraining as) what is reported in Ref.~\cite{Abbott:2017wau} for DES \yone.  When we split \om, the \omgrow-\aia degeneracy causes the DES-only constraints to widen considerably, with the bulk of the posterior volume residing in the region with small \omgrow and high \aia. Combining external data places a lower bound on \omgrow, which breaks the degeneracy and restricts \aia to small values. In fact, DES+Ext-geo and DES+Ext-all constraints on  \aia are slightly tighter and peak at slightly lower values in our split parameterization compared to \lcdm.  We observe a slight negative degeneracy between \aia and \mnu in these combined posteriors, so it is possible that this is caused by the same properties of the data which drive the high \mnu results.




\section{Discussion}\label{sec:conclusion}

We perform a combined analysis of DES Y1  galaxy clustering and weak lensing,
DES Y1 BAO, and DES Y3 supernova measurements in which we
split cosmological parameters related to the physics of dark energy into
separate ``growth'' and ``geometry'' versions. In this growth-geometry split
analysis, the geometry parameters \paramsgeo enter model predictions for
observables related to expansion history, including all distances, the shape
of the high-$z$ matter power spectrum, and projection operations used to
convert the three-dimensional power spectrum to observed 2PCF. The growth
parameters, in turn, enter calculations of
late-time structure growth:  \paramsgrow are used to compute the linear and non-linear  evolution of the matter power spectrum at late times as well as intrinsic alignment contributions to shear correlations.

We primarily focus on splitting \om, and our main results are reported based
on two data combinations:  DES+Ext-geo, which combines the DES measurements
with external geometric information from BOSS DR12 BAO and a compressed Planck
2015 likelihood, and DES+Ext-all, which additionally includes  BOSS DR12 RSD
measurements  as an external probe of structure growth. To supplement these
main results, we also consider secondary data-model combinations which are
less robust to changes in our modeling assumptions but can still aid in the
interpretation of the main results: DES-only constraints on split \om, and
DES+Ext-geo constraints when splitting both \om and $w$. We
  stress-test our analysis procedure by ensuring that the results are not
  biased in the presence of a sample of injected systematic errors, and
  perform a  blinded analysis; see Sec.~\ref{sec:validation} for details.

We use these analyses to address the questions raised in
  the introduction, which we now answer in order.

\subsection{Are DES constraints informed more by growth or geometric information?}

For all data combinations considered, we find constraints on geometric parameters to be much tighter than those on growth parameters (\fig{fig:sOm_results}). Thus, at least in the context of how we have  defined growth and geometric observables, DES constraints are more informed by  geometry than by a direct measurement of the evolution of structure growth. This is both  because changing \omgrow  has a smaller effect on the matter power spectrum than \omgeo (see Fig.~\ref{fig:PkvaryOm}), and because of parameter degeneracies impacting the \omgrow constraints.  As  seen in Fig.~\ref{fig:PkvaryOm}, changing \omgrow results in a nearly scale-independent amplitude change to $P(k)$. Consequently, \omgrow will be largely degenerate with any parameters that change the amplitude of the DES signal.  In the case of the lensing information, which contributes significantly to the overall constraints, the amplitude is controlled by $S_8$, and is largely degenerate with the neutrino mass and intrinsic alignment parameters.  Consequently, degeneracies between these parameters and \omgrow degrade the \omgrow constraints.

In fact, it is only when we combine two independent measures of structure growth, in the DES+Ext-all data combination, that we are  able to fully constrain \omgrow. For both DES+Ext-geo (where growth information comes only from DES) and Ext-all (where growth information comes only from BOSS RSD), we see that constraints on late-time structure growth are limited by our inability to break parameter degeneracies.
The Ext-all measurements do not constrain \omgrow on their own because of a  degeneracy between \omgrow and \mnu (\fig{fig:degeneracies_sOm_p5br}), 
while DES+Ext-geo is mainly limited by its inability to distinguish between \omgrow and \aia (\fig{fig:degeneracies_sOm_p5b}). When we combine all of these data together as DES+Ext-all, these degeneracies are broken and we are able to constrain \omgrow.
When we fix the sum of neutrino masses to
  0.06 eV, we find that either  Ext-all (with no DES data) or  DES+Ext-geo  are able to constrain both \omgeo and \omgrow on their own (\fig{fig:fixnu-da-sp5br-small}).

When we split both \om and $w$,  a
significant degeneracy between \omgrow and \wgrow prevents us from being able
to constrain the growth parameters even with DES+Ext-all data (\fig{fig:sOm-sw_results-bigplot}). This
suggests that additional growth probes would need to be included in order to
provide enough redundancy to distinguish between \om-like and $w$-like
deviations from standard structure growth.

\subsection{Are the data consistent with  \paramsgrow=\paramsgeo?}

Our constraints on $\paramsgrow$ and $\paramsgeo$ are statistically consistent, in the sense that we find tensions with \lcdm to be less than $2\sigma$ when assessed using either the marginalized posterior for the difference $\omgrow-\omgeo$, or  Bayesian Suspiciousness. 
For both the DES+Ext-geo and DES+Ext-all data combinations,  the
bulk of our posterior resides in the part of parameter space where
$\omgrow>\omgeo$ (\fig{fig:diff-sOm}). This preference   is not seen for DES data alone,
where  degeneracies with \as, and \aia prevent one from placing a lower bound on the growth parameter (\fig{fig:explaining-projection-effects}), shifting the posterior towards low \omgrow. Equating the fraction of the posterior volume above the \lcdm line of $\omgrow=\omgeo$ with equivalent one-sided $p$-values for a Gaussian distribution, 
  we find that the DES-only posterior is in  agreement with \lcdm at the $0.5\sigma$ level, while DES+Ext-geo and DES+Ext-all are  consistent with \lcdm at $1.3\sigma$ and $1.6\sigma$ levels, respectively.   Bayesian Suspiciousness quantifies the agreement of our posterior with $\paramsgrow=\paramsgeo$ in the model's full parameter space. According to this metric, when we split \om the DES-only posterior has a $1.5\sigma$ tension with \lcdm, DES+Ext-geo's tension is $1.9\sigma$ and DES+Ext-all's is $1.0\sigma$.  Fixing $\mnu=0.06$ eV brings all three data combinations into $<1\sigma$ agreement with \lcdm. When we split both \om and $w$ we find a DES-only tension with \wcdm of $1.6\sigma$ and a DES+Ext-all tension of $1.4\sigma$. 
We caution that given the strong degeneracy between \omgrow and \aia, these results will be sensitive to changes in our model of the redshift dependence of intrinsic alignments (\fig{fig:l-sOm_pipetests}).

\subsection{Is the DES preference for low \sigeight compared to Planck driven more by geometry or growth?}

This question is not straightforward to answer,
but our results support the idea that the \seight tension is driven  by constraints on the evolution of structure, as opposed to a mismatch between DES and Planck geometric constraints  which somehow propagates into the \seight parameter direction.  To explain this conjecture, we  note that for all  data  combinations we consider,  constraints on geometry parameters are very similar to   their un-split \lcdm or \wcdm constraints. This means that our split parameterization can be viewed as a generic way of allowing the properties of structure growth to vary around a fixed \lcdm background. Thus, if splitting growth and geometry absorbs the offset between DES and Planck \sigeight measurements into a deviation from $\omgrow=\omgeo$
, this would suggest that modifications to structure growth  explain the \sigeight tension. The question therefore becomes: does our split parameterization relieve tension between DES and Planck?

The datasets we consider somewhat complicate this assessment because we do not analyze the full Planck likelihood that is typically used to quantify the \seight tension. The closest comparison we can make is between DES and Ext-geo constraints. The  Ext-geo data are less constraining than typical  analyses of the full Planck likelihood with $A_{\rm Lens}=1$, 
enough so that even in \lcdm their $1\sigma$ confidence regions for  \sigeight overlap substantially with those from DES.  Since we construct the compressed Planck likelihood specifically to be insensitive to late-time structure growth (including by marginalizing  over $A_{\rm Lens}$) this could be a clue that  it is in fact growth-related  CMB observables that  drive the Planck preference for high \sigeight relative to DES.

That being said, our results may still contain some indication of the data properties which drive the \sigeight tension in \lcdm. As we discuss in \sect{sec:mnu_results},  our DES+Ext-geo, Ext-all, and DES+Ext-all constraints on \mnu depend sensitively on the relative value of \as measured by Planck compared to \seight measured by either DES or BOSS RSD (\fig{fig:growthrectangle_extal}). Thus, the fact that those constraints favor high neutrino mass --- and consequently, high \omgrow --- is possibly an indication that the CMB data prefer a higher density fluctuation amplitude  than the LSS observables.

On the topic of tensions, we additionally note that splitting growth and geometry has almost no impact on  $H_0$ constraints (\figs{fig:degeneracies_sOm_p5br},~\ref{fig:degeneracies_sOm_p5b}). This supports the idea that it is difficult to resolve the Hubble tension with simple model extensions to \lcdm which alter late-time structure growth, echoing arguments made in Refs.~\cite{Knox:2019rjx,Aylor:2018drw,Bernal:2016gxb}, as well as findings from studies of decaying dark matter~\cite{Chen:2020iwm,Clark:2020miy,Pandey:2019plg},  modified gravity, and coupled dark energy models~\cite{Ade:2015rim,Aghanim:2018eyx}.

\subsection{Comparison to previous results}\label{sec:resultscompare}

We now compare our results to those from other geometry-growth analyses in the
literature. Direct comparisons are made challenging by the fact that each work
made different choices in how to define the geometry-growth split, in addition to
using different datasets and applying different modeling of the
systematics. Nevertheless some general conclusions can nevertheless be drawn
from these comparisons.

Our modeling choices are closest to those in Ref.~\cite{Ruiz:2014hma}. They
combine the CFHTLens weak lensing with an early-Universe prior based on Planck
2013 data (which is somewhat comparable to our DES+Ext-geo data combination
which also has additional geometric constraints from supernovae and BAO), and
also include galaxy cluster abundances which are sensitive to both geometry
and growth. The fiducial analysis in Ref.~\cite{Ruiz:2014hma} is however less
conservative than ours as it fixes neutrino mass and does not include any
intrinsic alignments in the weak lensing modeling. As a result, even though
Ref.~\cite{Ruiz:2014hma} uses less constraining data, they constrain
\omgrow more tightly than we do; the strength of our \omgrow constraints
becomes comparable to theirs when we fix neutrino mass (see
\app{app:fixmnu}). These differences aside, they agree with us in finding
that \omgeo is better constrained than \omgrow, that the constraints are
compatible with $\omgrow=\omgeo$, and that the majority of the posterior
resides in the $\omgrow>\omgeo$ part of parameter space. When splitting both
$w$ and \om Ref.~\cite{Ruiz:2014hma} finds a $3\sigma$ preference for
$\wgrow>\wgeo$, strongly indicating less structure than would be expected given
constraints from expansion history, though letting \mnu vary entirely removes
that tension in favor of high neutrino masses.

References~\cite{Wang:2007fsa} and~\cite{Bernal:2015zom} perform a
``perturbations vs. background'' split and use growth parameters to compute
CMB anisotropy properties, rather than classifying growth parameters as
specific to {\em late-time} structure evolution as done in this paper.
Therefore the split-\om model in these references probes different physics and
is not directly comparable to our results. However, their results from
splitting only $w$ will be sensitive to only late-time growth-geometry
discrepancies, and are more similar to what we study. In this split-$w$-only
test, Refs.~\cite{Wang:2007fsa,Bernal:2015zom} find consistency with
\wcdm. When Ref.~\cite{Bernal:2015zom} splits both \om and $w$, they find
$\wgrow>\wgeo$ at $3.5\sigma$, in agreement with a similar analysis in
Ref.~\cite{Ruiz:2014hma}, indicating less structure seen by growth observables
than geometric ones. However because of the differences in the analyses,
datasets, and treatment of the systematics, we caution against overinterpretation of that comparison.

\subsection{Outlook}\label{sec:outlook}
The increasing precision of cosmological measurements will provide opportunities to perform more stringent tests of our standard cosmological model, including via future iterations of growth-geometry split analyses like the one presented here. In the coming months, updated Y3 galaxy clustering and weak lensing measurements will be released which have roughly three times the sky area, greater depth, and advances in methodology compared to Y1. Those measurements will provide improved constraints on both cosmological parameters and, crucially for testing growth-geometry consistency, the properties of intrinsic alignments.

It is worth noting that the external likelihoods in this paper were chosen to follow versions used in other DES Y1 papers~\cite{Abbott:2018xao}, updated versions of both the Planck and BOSS likelihoods are already available and could be easily applied to near-future growth-geometry split studies.
  Relative to Planck 2015, the Planck 2018 cosmology results~\cite{Aghanim:2018eyx} have slight shifts in several parameters that would affect our compressed likelihood. Of these, the most impactful is that Planck 2018's improved polarization measurements lead to constraints on \as which shift to lower values as they narrow by about a factor of two.  As we saw in \fig{fig:growthrectangle_extal},  even a small change in Planck's \as constraints can have a significant impact on the region of overlap  between CMB and late-time growth measurements in the \mnu-\sigeight plane. Lowering and tightening Planck's \as constraint may be enough to shift Ext-all towards favoring low rather than high values of neutrino mass. This would in turn likely lower the values of \omgrow preferred by both DES+Ext-geo and DES+Ext-all.

  One could also consider updating our BAO and RSD measurements to use the recently released  eBOSS DR16~\cite{Alam:2020sor} measurements, which combine the DR12 BAO galaxies we use with the low-$z$ Main Galaxy Sample (MGS) sample, high-$z$ eBOSS galaxies (LRG and ELG), high-$z$ quasars, and Lyman-$\alpha$ forest mesurements. The high-$z$ BAO measurements tend to  prefer lower values of \om and $h$ than the galaxy samples used in DR12, so it is likely that switching to eBOSS would pull our constraints to slightly lower \omgeo.  However, as our analysis is currently more limited by its ability to constrain growth rather than geometry, the largest impact of switching from BOSS DR12 to eBOSS would be from the inclusion of additional RSD measurements,
  from the MGS, ELG, LRG, and QSO samples. Nearly all of these  added samples have \fsig constraints that are high relative to the prediction from the Planck 2018 best-fit cosmology, so updating to eBOSS RSD  would likely raise the \sigeight value preferred by Ext-all. Referencing \figs{fig:degeneracies_sOm_p5br} and~\ref{fig:growthrectangle_extal}, we predict that this would likely  pull our DES+Ext-all results to lower \omgrow  by making the Ext-all posterior  more compatible with small neutrino mass. 

Beyond increasing precision of individual measurements, including additional, complementary probes of structure growth could benefit future growth-geometry split analyses. 
We found in this analysis  that including growth information from both DES and RSD allowed us to  break degeneracies between \omgrow, \mnu, and \aia in order to more robustly test \lcdm. Adding more observables that are sensitive to structure growth can help us further disentangle searches for deviations from \lcdm from the effects of neutrino mass or astrophysical systematics.  One approach to doing this could be to use full-shape information in the galaxy correlations measured by BOSS to directly constrain cosmological parameters, as in Refs~\cite{Troster:2019ean,DAmico:2019fhj,Ivanov:2019hqk}. 
Another, would be to include measurements of galaxy clusters.
Previous growth-geometry split analyses in Refs.~\cite{Ruiz:2014hma,Bernal:2015zom}  report that  galaxy cluster number counts significantly influence their growth parameter constraints, though that is complicated by systematics related to the calibration of mass-observable relations. Thus, combining galaxy clustering and weak lensing data with  galaxy cluster counts, as in Ref.~\cite{To:2020way}, may be a powerful way to break degeneracies with systematics and add constraining power to future tests of growth-geometry consistency. Finally, another promising avenue could be to include CMB lensing data in a combined analysis like those in Refs.~\cite{Abbott:2018ydy,Baxter:2018kap}. Since the CMB lensing kernel reaches higher redshifts than galaxy lensing, this would give us a longer line-of-sight lever arm for probing how LSS has evolved over time.

Looking further ahead, searching for deviations from the predictions of \lcdm, particularly in the evolution of structure growth, will be a core part of future cosmological experiments, including DESI, the  Rubin Observatory Legacy Survey of Space and Time, the Nancy Grace Roman Space Telescope, and Euclid, as well as the Simons Observatory and CMB S4.
These searches may be conducted in a variety of ways, using parameterizations that range from purely phenomenological splits of data to more physical models derived from modified-gravity actions. Whatever the approach, some findings 
of this growth-geometry split analysis  are  broadly applicable: as measurements get more precise, it will only become more important to characterize how searches from beyond-\lcdm physics are influenced by the assumptions about massive neutrinos and  astrophysical systematics like intrinsic alignments, and a key way to distinguish between those things will be by performing combined analyses of multiple probes of structure growth.

\section*{Acknowledgments}

This paper has gone through internal review by the DES collaboration. 
JM has been supported by the Porat Fellowship at Stanford University, and by the Rackham Graduate School through a Predoctoral Fellowship.

The analysis made use of the software tools  {\sc SciPy}~\cite{Jones:2001}, {\sc NumPy}~\cite{Oliphant:2006},  {\sc Matplotlib}~\cite{Hunter:2007}, {\sc CAMB}~\cite{Lewis:1999bs,Howlett:2012mh}, {\sc GetDist}~\cite{Lewis:2019}, {\sc Multinest}~\cite{Feroz:2007kg,Feroz:2008xx,Feroz:2013hea},  {\sc Polychord}~\cite{Handley:2015fda}, {\sc anesthetic}~\cite{Handley:2019mfs}, \cosmosis~\cite{Zuntz:2014csq}, and {\sc Cosmolike}~\cite{Krause:2016jvl}.
It was supported  through computational resources and
services provided by the National Energy Research Scientific Computing Center (NERSC), a U.S. Department of Energy Office of Science User Facility operated under Contract No. DE-AC02-05CH11231; and by the Sherlock cluster, supported by Stanford University and the Stanford Research Computing Center.

Funding for the DES Projects has been provided by the U.S. Department of Energy, the U.S. National Science Foundation, the Ministry of Science and Education of Spain, 
the Science and Technology Facilities Council of the United Kingdom, the Higher Education Funding Council for England, the National Center for Supercomputing 
Applications at the University of Illinois at Urbana-Champaign, the Kavli Institute of Cosmological Physics at the University of Chicago, 
the Center for Cosmology and Astro-Particle Physics at the Ohio State University,
the Mitchell Institute for Fundamental Physics and Astronomy at Texas A\&M University, Financiadora de Estudos e Projetos, 
Funda{\c c}{\~a}o Carlos Chagas Filho de Amparo {\`a} Pesquisa do Estado do Rio de Janeiro, Conselho Nacional de Desenvolvimento Cient{\'i}fico e Tecnol{\'o}gico and 
the Minist{\'e}rio da Ci{\^e}ncia, Tecnologia e Inova{\c c}{\~a}o, the Deutsche Forschungsgemeinschaft and the Collaborating Institutions in the Dark Energy Survey. 

The Collaborating Institutions are Argonne National Laboratory, the University of California at Santa Cruz, the University of Cambridge, Centro de Investigaciones Energ{\'e}ticas, 
Medioambientales y Tecnol{\'o}gicas-Madrid, the University of Chicago, University College London, the DES-Brazil Consortium, the University of Edinburgh, 
the Eidgen{\"o}ssische Technische Hochschule (ETH) Z{\"u}rich, 
Fermi National Accelerator Laboratory, the University of Illinois at Urbana-Champaign, the Institut de Ci{\`e}ncies de l'Espai (IEEC/CSIC), 
the Institut de F{\'i}sica d'Altes Energies, Lawrence Berkeley National Laboratory, the Ludwig-Maximilians Universit{\"a}t M{\"u}nchen and the associated Excellence Cluster Universe, 
the University of Michigan, NFS's NOIRLab, the University of Nottingham, The Ohio State University, the University of Pennsylvania, the University of Portsmouth, 
SLAC National Accelerator Laboratory, Stanford University, the University of Sussex, Texas A\&M University, and the OzDES Membership Consortium.

Based in part on observations at Cerro Tololo Inter-American Observatory at NSF's NOIRLab (NOIRLab Prop. ID 2012B-0001; PI: J. Frieman), which is managed by the Association of Universities for Research in Astronomy (AURA) under a cooperative agreement with the National Science Foundation.

The DES data management system is supported by the National Science Foundation under Grant Numbers AST-1138766 and AST-1536171.
The DES participants from Spanish institutions are partially supported by MICINN under grants ESP2017-89838, PGC2018-094773, PGC2018-102021, SEV-2016-0588, SEV-2016-0597, and MDM-2015-0509, some of which include ERDF funds from the European Union. IFAE is partially funded by the CERCA program of the Generalitat de Catalunya.
Research leading to these results has received funding from the European Research
Council under the European Union's Seventh Framework Program (FP7/2007-2013) including ERC grant agreements 240672, 291329, and 306478.
We  acknowledge support from the Brazilian Instituto Nacional de Ci\^encia
e Tecnologia (INCT) do e-Universo (CNPq grant 465376/2014-2).

This manuscript has been authored by Fermi Research Alliance, LLC under Contract No. DE-AC02-07CH11359 with the U.S. Department of Energy, Office of Science, Office of High Energy Physics.

\appendix

\section{The impact of changing $z_i$}\label{sec:varyzi}
\begin{figure}
  \includegraphics[width=.8\linewidth]{{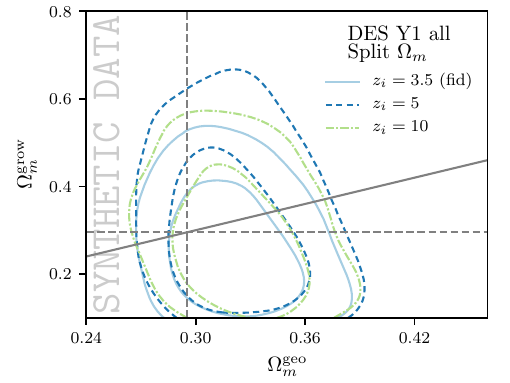}}\\
  \includegraphics[width=.8\linewidth]{{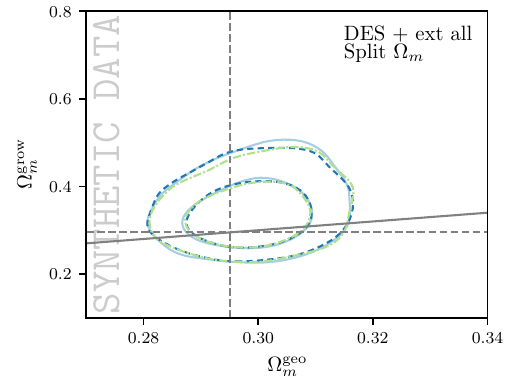}}
\caption{The impact of changing the redshift $z_i$ where growth parameters start controlling the evolution of the matter power spectrum, defined in \eq{eq:mixpower}, for DES-only (top panel) and DES+Ext-all (bottom panel). Countours show the 68\% and 95\% confidence regions from the analysis of synthetic data.}
\label{fig:zi_impact}
\end{figure}

Here we   characterize how our expected constraints depend on $z_i$, the redshift from \eq{eq:mixpower} below which the evolution of the linear matter power spectrum is controlled by growth parameters. In addition to our fiducial choice of $z_i=3.5$, we analyze the fiducial synthetic data vector with versions of our pipeline that have $z_i=5$ and $z_i=10$.

\fig{fig:zi_impact} shows the impact of these changes.
For DES data alone we find that increasing $z_i$ from 3.5 to 5 or 10 does slightly affect the upper bounds on \omgrow, but does not significantly impact constraints on \omgeo. In this simulated analysis we find that $z_i=5$ results in the weakest \omgrow upper bound, while the bound from  $z_i=10$ is similar to but slightly weaker than the fiducial  $z_i=3.5$ result. The reason for the lack of monotonic trend in this is not clear, but given DES-only's poor constraining power on  growth parameters and its sensitivity to projection effects and systematics (to be discussed in subsequent appendices), we refrain from over interpreting this. It is also possible that some of this variation is due to noise in the  posterior estimate from Multinest, which can occur because of the small number of samples the posterior tails.  Results from the joint analysis of DES+Ext-all are not significantly affected by changing $z_i$.

\section{Evaluating projection effects}\label{sec:syndat}

This Appendix discusses insights gained by studying the results of a simulated analysis in which we applied our parameter estimation pipeline to noiseless synthetic data generated by setting observables (e.g. BAO $\alpha$ parameters, weak lensing 2PCF) equal to theory predictions at fiducial parameter values. Specifically here we focus on the case where the  synthetic data is generated using our baseline pipeline -- that is, using the same modeling choices as in the theory predictions that we employ for for parameter estimation.
Before analyzing real data, we used these simulated analyses to identify data and model combinations for which we expect  marginalized posteriors to be reliably informative about whether growth and geometry constraints are consistent. This is not guaranteed: in high-dimensional parameter spaces like the one we consider,
projecting the posterior volume onto one- or two-dimensional subspaces can result in offsets between the peaks of marginalized posteriors and the best fit parameter values. By comparing marginalized posteriors from simulated analyses to the known input values, we are able to characterize the extent of these projection effects on our split parameters.  We use this comparison to identify which data combinations will the focus of our analysis.

Our main results will be derived from the model and data combinations whose marginalized posteriors in simulated analyses  are consistent with the input parameter values. These are:
\begin{itemize}
\item  Split \om constrained with DES+Ext-geo, and
\item Split \om constrained with DES+Ext-all.
\end{itemize}
For both of these model-data combinations,
the input parameter values are contained within 68\% confidence contour for the synthetic-data version of all two-dimensional constraint plots appearing in this paper. The input values also are within the 68\% confidence interval of the one-dimensional marginalized posterior of the split parameter, as well as their differences.

We also consider two additional sets of constraints,
\begin{itemize}
\item Split \om constrained by DES only, and
\item Split \om and $w$ constrained by DES+Ext-all,
\end{itemize}
for which we find offsets between the input parameter values and the peaks of the marginalized 1D posteriors of the split parameters.
In our simulated analysis, the DES-only marginalized posteriors for \omgeo and \omgrow are biased (high and low, respectively) relative to their input values by about $1\sigma$. The DES+Ext-all constraints on split \om and $w$ exhibit $\sim 1\sigma$ offsets for marginalized posteriors of \omgrow and \wgrow.
We therefore treat the results from these constraints with caution.
Because we do not trust the one-dimensional posterior peaks to accurately reflect the best-fit values, we will not quote their one-dimensional marginalized parameter constraints.
However, we will still report model-comparison measures
and will  show constraint contours for two-dimensional marginalized posteriors. This is  motivated by the fact that in our simulated analyses the  68\% confidence intervals of these two-dimensional marginalized posteriors do contain the input parameter values.
Since simulated analysis results for DES+Ext-all, our most constraining dataset, results in constraints on split \om and $w$ that are offset from their input values, we do not consider constraints on the split $w$  model from the less constraining data combinations, DES-only and DES+Ext-geo.

  \begin{figure}
  \centering
    \includegraphics[width=\linewidth]{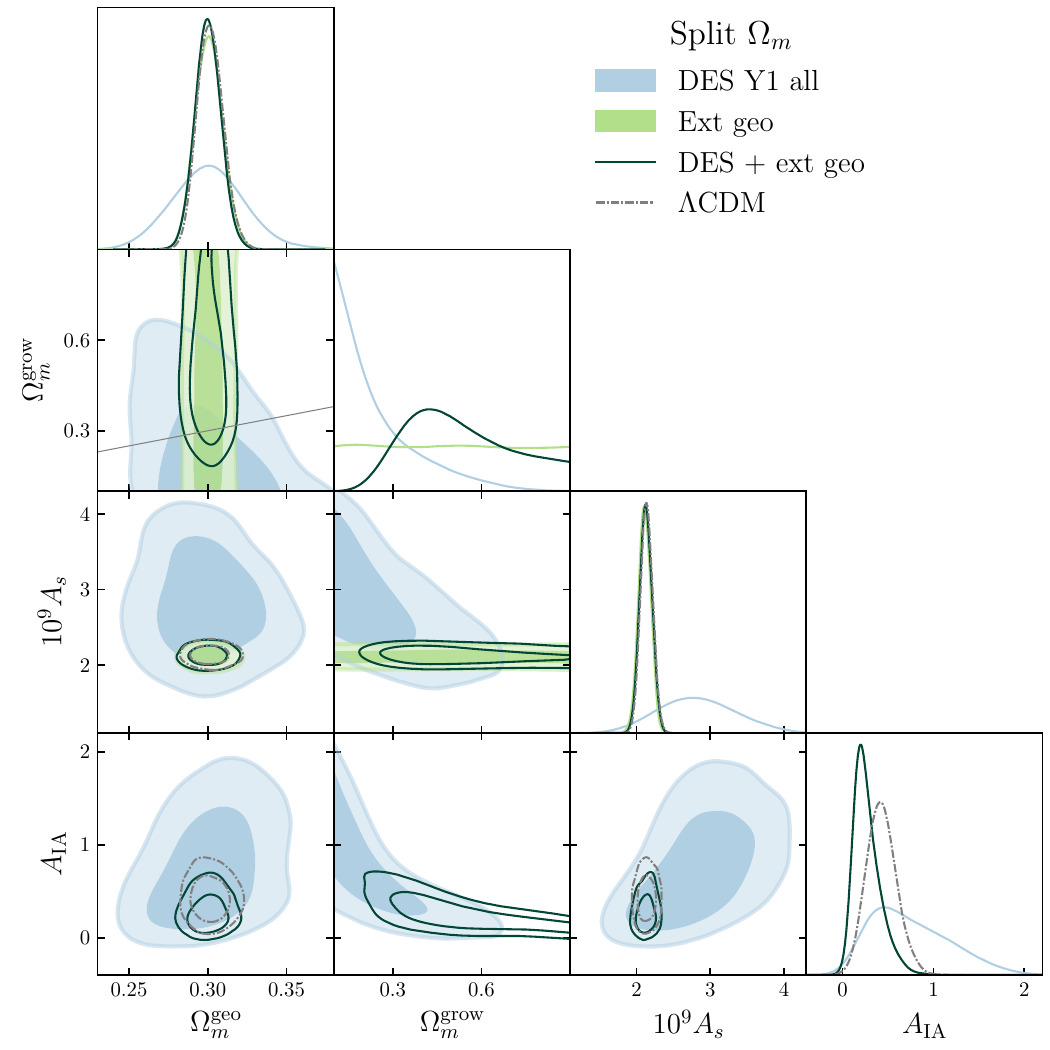}
    \caption{Constraints  illustrating the parameter degeneracies that are relevant to understanding the parameter space projection effects impacting the DES-only constraints on split \om. Off diagonal panels show 68\% and 95\% confidence intervals, with DES-only results in blue, Ext-geo in light green, and DES+Ext-geo as the dark green unshaded contours. The gray diagonal line shows where $\omgrow=\omgeo$, and gray dashed contours show \lcdm results for DES+Ext-geo.
    }
  \label{fig:explaining-projection-effects}
\end{figure}

It can be instructive to examine the parameter degeneracies that drive the projection effects described above.  The fact that the DES-only constraints on split \om are biased high for \omgeo and low for \omgrow  can be understood in terms of a degeneracy between \omgrow and \omgeo, as well as degeneracies \omgrow has with the primordial power amplitude \as and with the intrinsic alignment amplitude \aia.
These degeneracies are illustrated (for real data) in \fig{fig:explaining-projection-effects}. Focusing initially on the DES-only constraints, we note that very low \omgrow values are allowed because they can be compensated by raising \as, while very high \omgrow values would presumably be ruled out based on the rate of structure growth occurring over the redshift range probed by DES.
  The degeneracy with \aia occurs because intrinsic alignment contributions to the shear 2PCF appear in \eq{eq:iadef} via a factor $\propto \aia\omgrow$. This causes the constant-posterior contours to have a banana shape in the \omgrow-\aia plane, such that small values of \omgrow allow large values of \aia and vice versa.  These  degeneracies combine with the fact that the DES-only likelihood is relatively flat in \omgrow (as can be seen in the profile likelihood shown in \fig{fig:iabins-likeprofile} below), to produce an  offset in the projected posterior.
  This translates into an offset in \omgeo as well because
    there is a weak degeneracy between \omgrow and \omgeo.  The DES+Ext-geo and DES+Ext-all constraints do not show these offsets because the Planck constraints  break the \omgrow-\as degeneracy.

The projection effects for DES+Ext-all constraints on split \om and $w$  are driven by the fact that the effects of \omgrow and \wgrow on observables are very degenerate with one another (see \fig{fig:sOm-sw_results-bigplot}). Though each of these growth parameters would have unbiased marginalized constraints if the other were fixed to its fiducial value, they are unconstrained when varied simultaneously. In other words,  while the data we consider can constrain deviations from standard structure growth, they are not informative enough to distinguish between \om-like and $w$-like changes.


\section{Impact of unmodeled systematics}\label{app:systests}

\begin{figure}
  \centering
  \includegraphics[width=\linewidth]{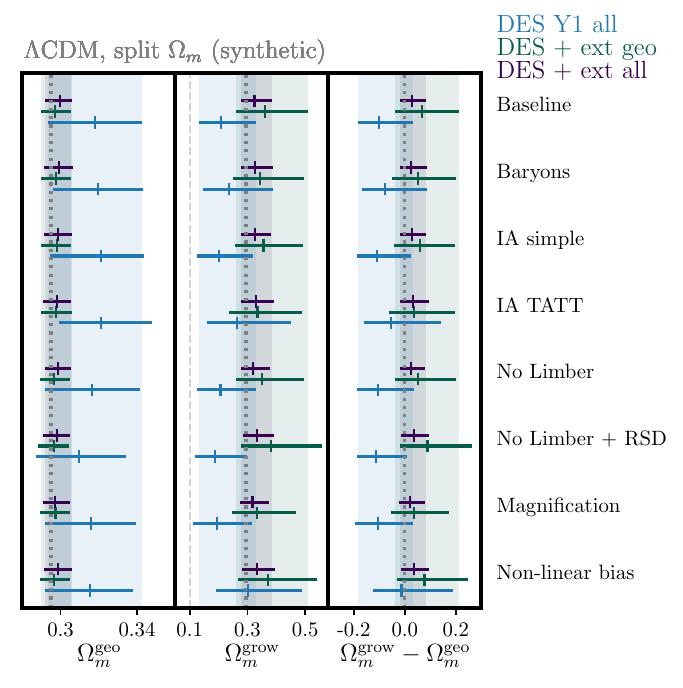}
  \caption{Robustness of constraints to adding systematics to simulated data, for the split-\om model. Data points and error bars represent the peak and 68\% confidence intervals for  marginalized one-dimensional posteriors. The vertical shaded regions correspond to the baseline  error bars.}
  \label{fig:l-sOm_systests}
\end{figure}


\begin{figure*}
\includegraphics[width=.5\linewidth]{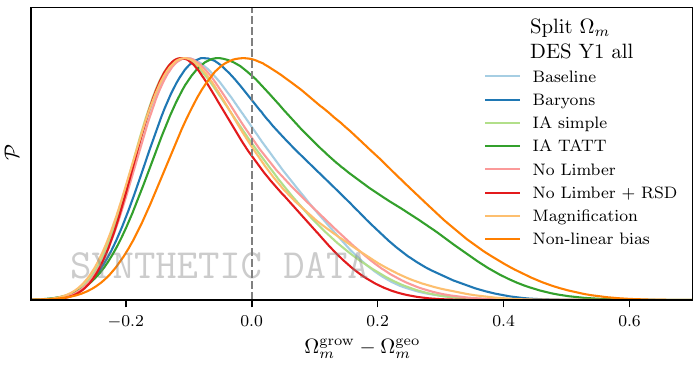}
\includegraphics[width=.25\linewidth]{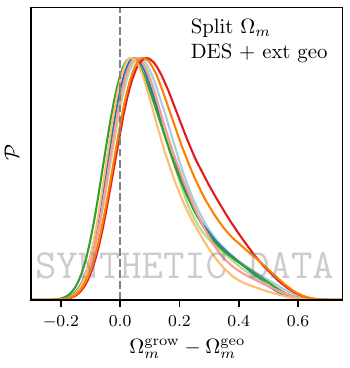}\\
\includegraphics[width=.25\linewidth]{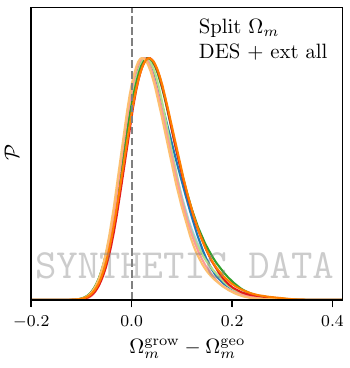}
\includegraphics[width=.25\linewidth]{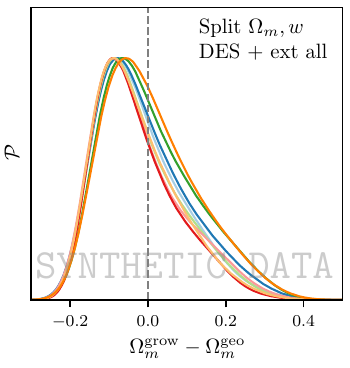}
\includegraphics[width=.25\linewidth]{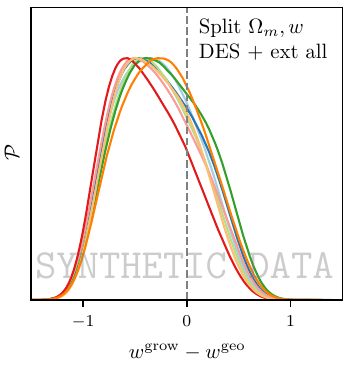}
\caption{
  Marginalized posteriors for synthetic data vectors contaminated by systematic effects, showing constraints on the difference between growth and geometry parameters. Top row: split-\om results for DES-only  (left) and for DES+Ext-geo (right).  Bottom row: Results from DES+Ext-all when we split \om (left), and when we split both \om and $w$, showing growth-geometry differences for \om (center) and $w$ (right).  }
  \label{fig:systests_differences}
\end{figure*}

We additionally analyze synthetic data where the DES \mpp  measurements are contaminated by unmodeled systematic effects in order to characterize our robustness against certain modeling assumptions.
For example, we compute the
\mpp observables using a non-linear galaxy bias model. By  treating those synthetic observables as if they are data and fitting with our fiducial model (which assumes linear galaxy bias),
we can quantify the extent to which unmodeled effects (here, the presence non-linear galaxy bias)
biases our cosmological results.
The synthetic data vectors  we use in this study are the same as those  used for similar tests in Refs.~\cite{Krause:2016jvl} and~\cite{Abbott:2018xao}. They are:
\begin{itemize}
\item Baseline - This data vector is equal to a theory prediction at a fiducial \lcdm cosmology, using the same modeling choices as parameter estimation.
\item Baryons - This data vector includes one case of possible effects of baryonic physics, the impact of AGN feedback on the non-linear power spectrum is included using the OWLS AGN hydrodynamical simulation~\cite{Schaye:2009bt}, following the method described in~\cite{Troxel:2017xyo}.
\item IA simple - Using the same nonlinear alignment model as in our fiducial model, this data vector is generated with $\aia=0.5$ and $\eta_{IA}=0.5$. We note that these parameters are marginalized over in our analysis, so including this data in our tests checks whether degeneracies between the intrinsic alignment parameters and the cosmological parameters can introduce biases.
\item IA TATT - Here,  the data vector is simulated with a different intrinsic alignment power spectrum shape. It is  modeled  by assuming all intrinsic alignments are generated by tidal torquing, which is quadratic in the tidal field, instead of the the linear tidal alignments described in our fiducial model.  To compute it, we use the Tidal Alignment and Tidal Torquing model (TATT)~\cite{Blazek:2017wbz} with tidal alignment amplitude $A_1=0$, tidal torquing amplitude $A_2=2$, and no $z$ dependence. 
\item No Limber - This data vector has been simulated using a theory calculation done without the Limber approximation for $w(\theta)$.
\item No Limber + RSD -  This data vector has been simulated using a theory calculation done without the Limber approximation and including the contributions of redshift space distortions for $w(\theta)$ as described in~\cite{Padmanabhan:2006cia}
\item Magnification - This data vector is simulated including contributions from magnification to $\gamma_t$ and $w(\theta)$, which are added in Fourier space as is described in~\cite{Bernstein:2008aq}.
\item Non-linear bias - This data vector goes beyond our fiducial model of linear galaxy bias and models the relationship between matter $\delta$ and galaxy density fluctuations $\delta_g$ as~\cite{McDonald:2009dh,Baldauf:2012hs}
  \lneqb\begin{equation}
    \delta_g = b_1^i\delta + \frac{1}{2}b_2^i[\delta^2-\sigma^2]
  \end{equation}\lneqe
where $\sigma$ is the variance in $\delta$ and $i$ refers to the lens redshift bin. This theory data vector was computed using the {\sc FAST-PT} code~\cite{McEwen:2016fjn} with input values  $b^i = \{1.45,1.55,1.65,1.8,2.0\}$ and the $b_2$ values used are estimated from fits to the Buzzard simulations~\cite{DeRose:2019ewy} to be $b_2 = 0.412 - 2.143 b_1 + 0.929 b_1^2 + 0.008 b_1^3$.
\end{itemize}
More detailed descriptions of the generation of these data can be found in Refs.~\cite{Krause:2016jvl} and~\cite{Abbott:2018xao}.

The metric for passing these tests is based on the one-dimensional marginalized posteriors for \omgrow and \omgeo, as well the \wgrow and \wgeo for the parameterization where $w$ is split.
For each of the data combinations discussed above, we verify that the  shift in the peak of the posterior is less than $0.3\sigma$ relative to the baseline analysis. We evaluate the size of these shifts by computing an effective $\sigma$ by summing the two posteriors' asymmetric 68\% confidence intervals in quadrature. To state this more specifically, let $\hat\theta$ be the one-dimensional marginalized posterior peak on parameter $\theta$, and suppose the baseline and contaminated constraints are labeled $A$ and $B$ such that $\hat\theta_A>\hat\theta_B$.
Additionally let $\theta_A^{\rm \tt low68}$ be the lower bound of the 68\% confidence interval for dataset $A$ and $\theta_B^{\rm \tt up68}$ be the upper bound of the 68\% confidence interval for dataset $B$. We quantify the size of the posterior shift as
\lneqb\begin{equation}\label{eq:peakshift}
\Delta_{\theta}= \frac{\hat\theta_A - \hat\theta_B}{\sqrt{(\hat\theta_a - \theta_A^{\rm \tt low68})^2 + (\theta_B^{\rm \tt up68} - \hat\theta_B)^2}}.
\end{equation}\lneqe

Summary plots showing the results for these tests are shown in \fig{fig:l-sOm_systests} for split \om and   \fig{fig:sOm-sw_systests} for split \om and $w$.
We additionally show in \fig{fig:systests_differences} how the posteriors from these same synthetic-data analyses project onto the one-dimensional marginalized posteriors of the differences $\omgrow-\omgeo$ and $\wgrow-\wgeo$.

Both of the main model and data combinations  identified in our fiducial simulated analysis of \app{sec:syndat} (DES+Ext-geo and DES+Ext-all constraints on split \om) pass these tests, as none of these changes result in a parameter shift larger than $0.3\sigma$. For split \om DES+Ext-geo results, the largest posterior shift observed is in \omgrow and occur when we add the effects of magnification to the synthetic data. The size of this shift is  $-0.16\sigma$ relative to the baseline simulated analysis.  For  DES+Ext-all, the systematic with the largest impact is non-linear galaxy bias, which shifts the \omgrow posterior by $+0.13\sigma$.

In addition to the prior volume effects described in \sect{sec:syndat}, the DES-only split \om constraints and the DES+Ext-all constraints on split \om and $w$ should be treated with caution because they fail these tests, in the sense that the some of the systematics produced parameter shifts larger than our desired $0.3\sigma$ threshold. For the DES-only split-\om results, this occurs for TATT intrinsic alignments, which changes the best-fit \omgrow by $+0.34\sigma$, and for nonlinear bias, which changes \omgrow by  $+0.51\sigma$. All other shifts are below $0.3\sigma$. For the DES+Ext-all constraints on split \om and $w$ the only systematic that generates a parameter shift larger than our threshold is the non-Limber and RSD modeling for galaxy clustering, which changes \wgrow by $-0.36\sigma$.

\section{Impact of changing analysis choices}\label{app:realdattests}

\begin{figure}
  \centering
  \includegraphics[width=\linewidth]{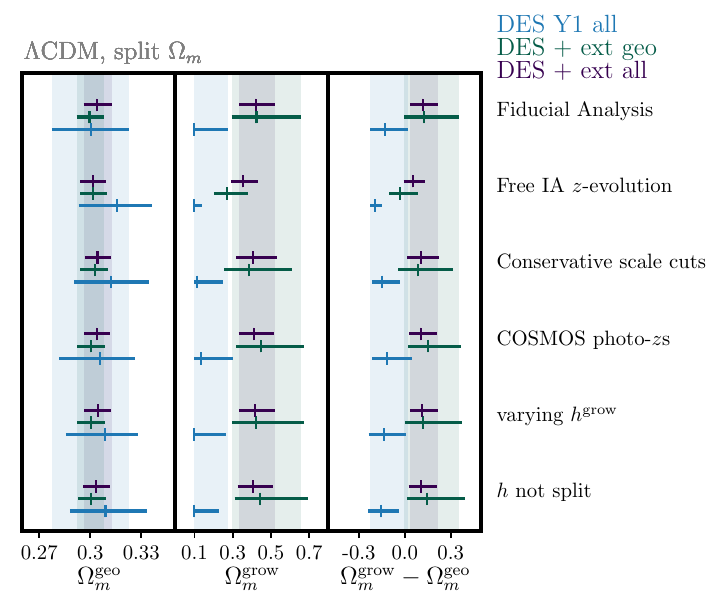}
  \caption{Robustness of real-data constraints  to changes in analysis choices when we split \om. Data points and error bars represent the peak and 68\% confidence intervals for  marginalized one-dimensional posteriors.  The vertical shaded regions correspond to the 68\% confidence interval of the  baseline measurements. }
  \label{fig:l-sOm_pipetests}
\end{figure}

\begin{figure*}
  \centering
  \includegraphics[width=.8\linewidth]{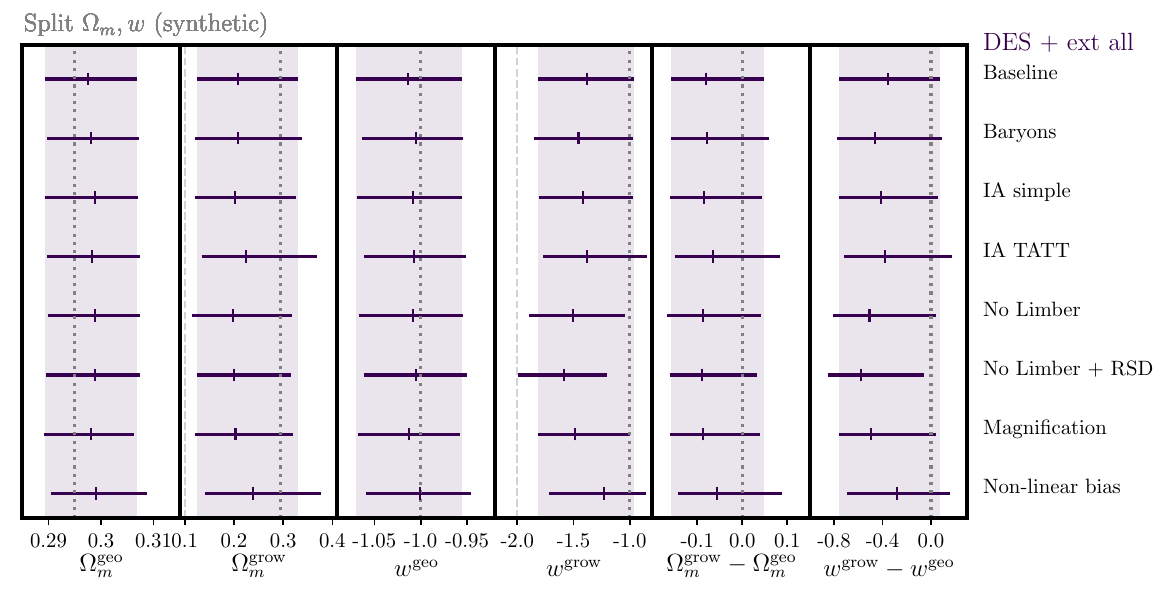}
  \caption{Robustness of results to adding systematics to simulated data for the model splitting \om and $w$. Data points and error bars represent the peak and 68\% confidence intervals for  marginalized one-dimensional posteriors. The vertical shaded regions correspond to the baseline  error bars.}
  \label{fig:sOm-sw_systests}
\end{figure*}
\begin{figure*}
  \centering
  \includegraphics[width=0.8\linewidth]{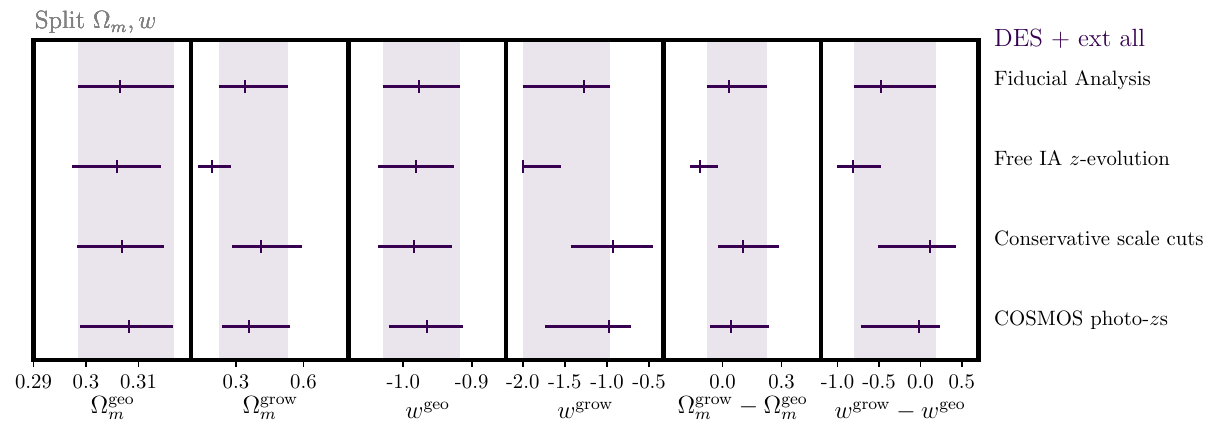} 
  \caption{Robustness of real-data constraints  to changes in analysis choices when we split \om and $w$. Data points and error bars represent the peak and 68\% confidence intervals for  marginalized one-dimensional posteriors.  The vertical shaded regions correspond to the 68\% confidence interval of the  baseline measurements.}
  \label{fig:sOm-sw_pipetests}
\end{figure*}

Before revealing the non-offset (unblinded) parameter estimates, we test the robustness of our analysis of real data against changes to various analysis choices. We perform this test similarly to the systematics tests described in \app{app:systests}, but instead of analyzing contaminated synthetic data, we compare parameter estimates obtained by running on the same real data, but altering aspects of our analysis pipeline. These changes are, following the analysis in Ref.~\cite{Abbott:2018xao}:
\begin{itemize}
\item Free IA $z$-evolution - Instead of assuming that the intrinsic alignment amplitude scales as a power law in redshift, we allow its amplitude to vary for independently for each source redshift bin.
\item Conservative scale cuts - We restrict our analysis to DES 2PCF measurements to large angles which are insensitive to non-linear LSS modeling.
\item COSMOS photo-$z$'s - We use alternative photometric redshifts for the DES source galaxies,  obtained by resampling  COSMOS data following the procedure in~\cite{Hoyle:2017mee}.
\end{itemize}
Additionally, for the split \om model we also show the results of additional tests examining the impact of changing the treatment of the Hubble parameter in our split parameterization. Recalling  our fiducial analysis splits $h$ and fixes $h^{\rm grow}=0.6881$, we show how parameter constraints change for:
\begin{itemize}
\item Varying $h^{\rm grow}$ - We allow $h^{\rm grow}$ to vary over the same $[0.1,0.9]$ prior range as $h\equiv h^{\rm geo}$.
\item $h$ not split - We require $h^{\rm grow}=h^{\rm geo}$ and vary it as in \lcdm.
\end{itemize}
These $h$ tests were conducted after the true analysis results were revealed (after unblinding).

The results of these  tests are summarized in \fig{fig:l-sOm_pipetests} for the split \om analysis, and in \fig{fig:sOm-sw_pipetests} for split \om and $w$. We  quantify the changes from the baseline analysis following the same method as in \app{app:systests} above.

Notably, for all data and model combinations  we see significant  parameter shifts in growth parameter estimates when we allow the intrinsic alignment amplitude to vary independently in each redshift bin.
For all pipeline variations other than free IA $z$-evolution, we find that our main results, DES+Ext-geo and DES+Ext-all constraints on split \om, are robust. For DES+Ext-geo, the largest parameter shift relative to the baseline analysis is a $+0.25\sigma$ change in \omgeo, which occurs when we switch to conservative scale cuts. For DES+Ext-all, the conservative scale cuts and not splitting $h$ tie for the largest shift, a $-0.10\sigma$ change in \omgrow.

The DES-only split \om results and the DES+Ext-all results for split \om and $w$ are less robust, even setting the free IA $z$-evolution results aside. For the DES-only constraints on split \om, conservative scale cuts produce a $0.38\sigma$ shift in \omgeo, while all other parameter shifts are below $0.3\sigma$. For DES+Ext-all constraints on split \om and $w$, using conservative scale cuts moves \omgrow by $0.31\sigma$ shift  and \wgrow by $+0.59\sigma$, and using the COSMOS photo-$z$s causes \wgrow to change by $0.36\sigma$.

After observing this behavior in parameter-offset (blinded) results from real data, we performed an analysis of synthetic data using the binned-IA model in order to better characterize its impact, and found that free IA $z$ evolution produced a similar change in posteriors.  We hypothesize that the large parameter shifts, especially in \omgrow, occurs when we introduce more freedom in IA redshift evolution  because of a parameter-space projection effect.
As  discussed for our fiducial NLA IA model in  \app{sec:syndat}, the fact that \omgrow is poorly constrained and degenerate with $\aia$ causes the  DES-only posterior to be skewed towards low \omgrow values. When we allow the IA amplitude to vary independently for each source redshift bin, this opens a large volume of parameter space where small \omgrow can compensate large IA amplitudes. That low-\omgrow posterior volume is much larger than the allowed region of parameter space  where \omgrow is high but all four IA amplitudes are small. This means that in the absence of strong constraints on \omgrow, small \omgrow values will dominate  one- or two-dimensional projections of the posterior. Degeneracies between \omgrow and other parameters will propagate that effect to other parameters like \omgeo. This is perhaps analogous to Ref.~\cite{Samuroff:2018xuo}'s finding that opening up ``too much'' freedom in the IA model causes \seight constraints to shift to smaller values, and we posit that this is why opening up additional IA parameter space causes such dramatic parameter shifts in \figs{fig:l-sOm_pipetests} and~\ref{fig:sOm-sw_pipetests}.

\begin{figure}
  \centering
  \includegraphics[width=.7\linewidth]{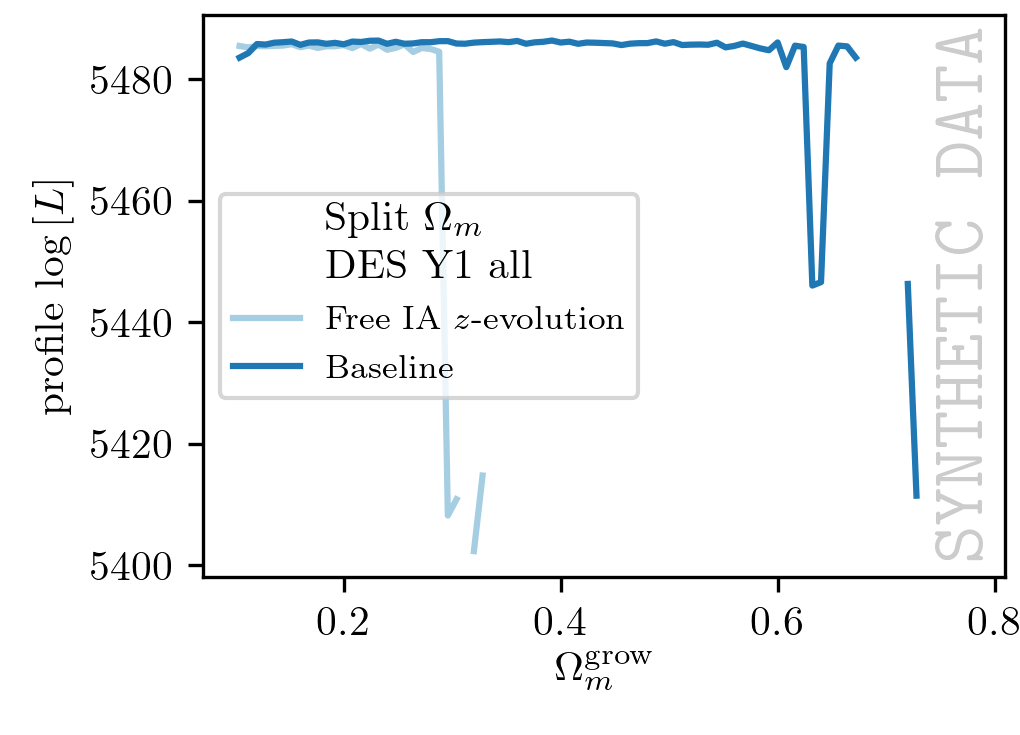}
  \includegraphics[width=.7\linewidth]{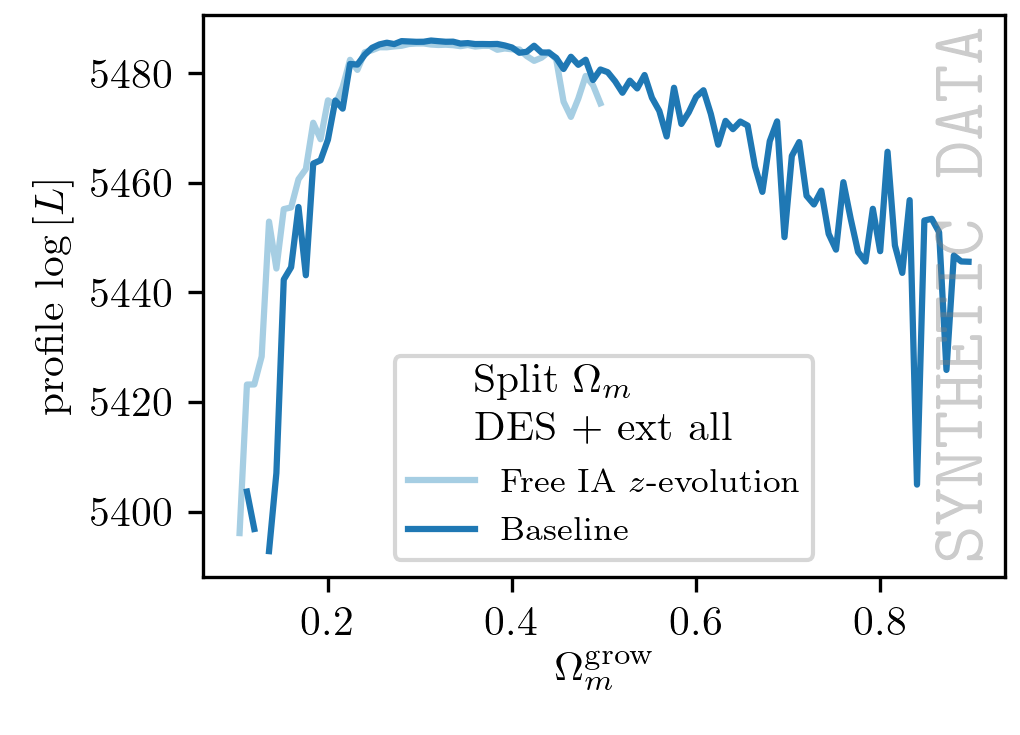} 
  \caption{Likelihood profile showing the maximum likelihood for chain samples within narrow bins of \omgrow. The sharp step functions in the DES-only plot show where the sample density decreases significantly due to the \aia-\omgrow projection effects discussed in \app{sec:syndat}. }
  \label{fig:iabins-likeprofile}
\end{figure}

To support this hypothesis, in \fig{fig:iabins-likeprofile} we show the profile likelihood for \omgrow for DES-only and DES+Ext-all constraints on synthetic data. The vertical axes of these plots show the maximum likelihood in our chain samples which have \omgrow within a narrow bin. These profiles are noisy and exhibit sharp drop-offs  because our sampler (Multinest in this case) returned very few chain samples in that region of parameter space.  Where there are enough samples to compare the baseline and binned-IA profiles, we see that they have  very similar profile likelihoods. This means that the ``free IA $z$-evolution'' model does not actually produce an improved fit to the data at small \om compared to our fiducial model. Rather, the posterior peaks at smaller \omgrow because binning IA increases the relative volume of the parameter space, and thus the density of chain samples, associated with small \omgrow compared to high \omgrow.   In other words, the change in posterior peak comes from parameter volume projection effects.

Current data~\cite{Samuroff:2018xuo,Abbott:2017wau,Abbott:2018ydy} are not able to rule out models with this much variation in the IA amplitude, but neither do they provide evidence for IA redshift evolution beyond our NLA power law, nor is there a strong theoretical motivation for it. Given this, we proceed with our analysis despite the nominal failure of this robustness test.

\section{Additional results}\label{sec:moreplots}

\subsection{Parameter degeneracies without RSD data}\label{sec:results_noRSD}

\begin{figure*}
  \centering
  \includegraphics[width=.6\linewidth]{{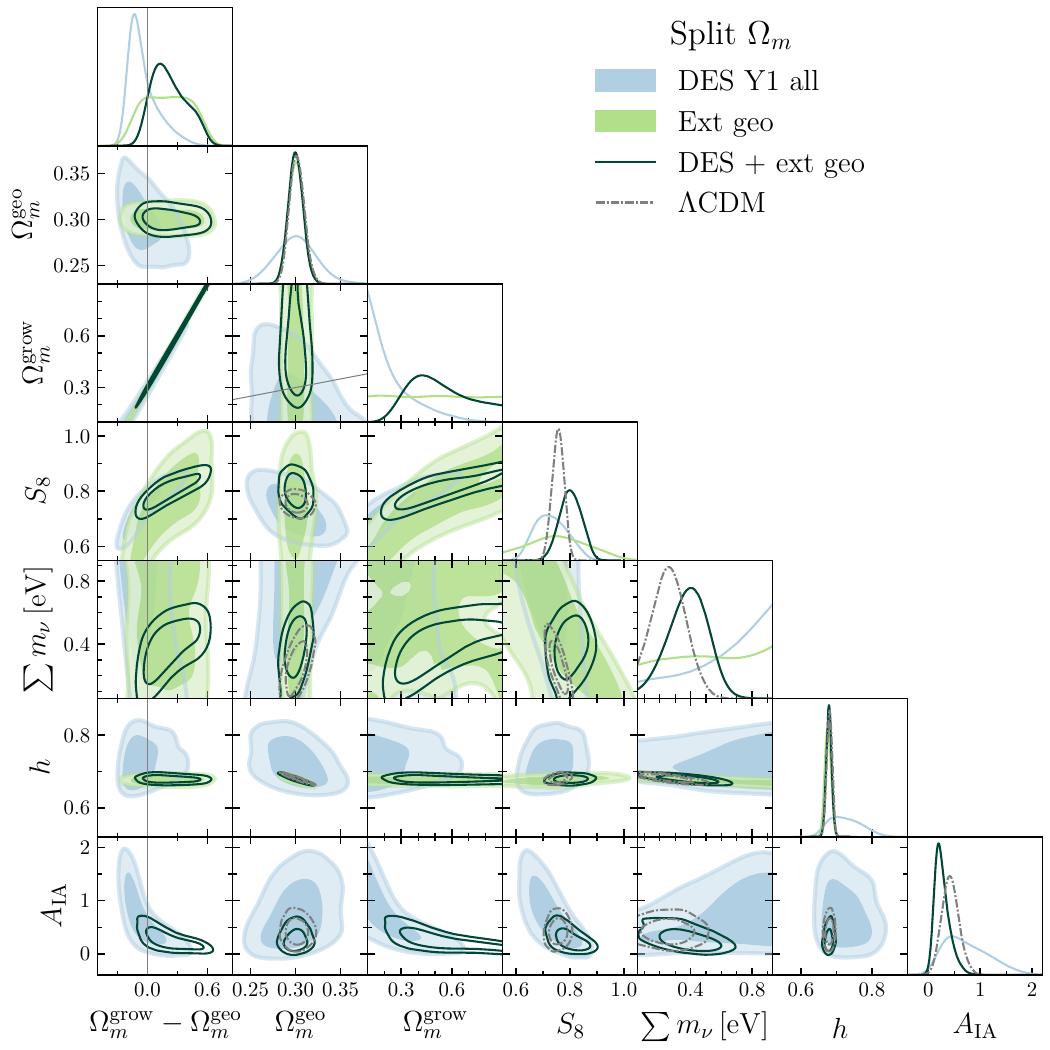}}
\caption{Marginalized constraints from DES and Ext-geo data. This plot is identical to  \fig{fig:degeneracies_sOm_p5br}, but uses the  external dataset that does not include BOSS RSD. }
\label{fig:degeneracies_sOm_p5b}
\end{figure*}

We include \fig{fig:degeneracies_sOm_p5b} to supplement the discussion  in \sect{sec:bigplot_results} about how splitting \om impacts constraints on \mnu, \seight, $h$ and \aia. It is identical to  \fig{fig:degeneracies_sOm_p5br}, except in that it does not include BOSS RSD measurements.  The off-diagonal panels show the 68\% and 95\% confidence intervals for DES-only  as blue shaded regions (identical to those in \fig{fig:degeneracies_sOm_p5br}), for Ext-geo in light green shaded regions, and for the DES+Ext-geo combination as dark green unshaded contours. Diagonal gray lines denote where $\omgrow=\omgeo$, and gray dashed contours show \lcdm results for DES+Ext-geo.  The diagonal panels show normalized marginalized posteriors for individual parameters.

\subsection{Impact of fixing \mnu}\label{app:fixmnu}

Here we present  additional information about the impact of fixing neutrino mass to supplement the discussion and \fig{fig:fixnu-da-sp5br-small} in \sect{sec:mnu_results}. \tab{tab:results_fixmnu} reports tension and model comparison metrics for the fixed neutrino mass analyses. It follows the same notation and conventions as what is used for the main results in \tab{tab:results}.

\begin{table}
  \begin{center}
    \caption{Constraints, tension, and model comparison statistics for split parameters when we fix $\mnu=0.06$ eV. }
    \label{tab:results_fixmnu}
    \renewcommand{\arraystretch}{1.3} 
    \begin{tabular}{c|ccc}

      \textbf{\lcdm,} & &    &   \\
      \textbf{ fix $m_{\nu}$} & DES &  DES+Ext-geo & DES+Ext-all  \\\hline
      \om  &  $0.279^{+0.023}_{-0.017}$ & $0.289^{+0.007}_{-0.005}$ & $0.290^{+0.007}_{-0.006}$\\
      $\tilde{d}$  & $ 13.4\pm 0.7 $ & $15.7\pm 0.7$ & $16.1\pm 0.7$ \\\hline

      $\log S^{\rm dat}$ & - & $-0.7\pm 0.2$ &  $ -1.1 \pm 0.2 $ \\
      $p(S>S^{\rm dat})$ & - &  $0.23 \pm 0.05$ &  $ 0.15 \pm 0.05 $ \\
      equiv. $\sigma$ & - & $1.2\pm 0.1$  &  $ 1.4 \pm 0.2$ \\\hline
      \multicolumn{4}{c}{}\\

      \textbf{Split \om,} &   & &  \\
      \textbf{fix $m_{\nu}$} & DES &  DES+Ext-geo &  DES+Ext-all  \\\hline
      $\omgrow - \omgeo$ &  -  & $-0.040^{+0.074}_{-0.055}$ & $-0.007^{+0.036}_{-0.037}$ \\
      \omgeo & - & $0.291^{+0.007}_{-0.007}$  &$0.292^{+0.006}_{-0.006}$\\
      \omgrow &  -  & $0.252^{+0.070}_{-0.051}$  & $0.284^{+0.036}_{-0.035}$\\
      $\tilde{d}$  & $ 14.5\pm 0.7 $& $ 17.6\pm0.9$ &$16.8\pm 0.8$ \\\hline

      $\log S^{\rm dat}$ & - & $-1.2\pm 0.3$ & $ -0.5 \pm 0.3$ \\
      $p(S>S^{\rm dat})$ &  - & $0.11 \pm 0.06$ &$0.29 \pm 0.06 $ \\
      equiv. $\sigma$ & -  & $1.6 \pm 0.3$ &$1.1 \pm 0.1$ \\\hline

      $\log S^{\rm mod}$ & $ 0.0 \pm 0.2 $& $0.4\pm 0.2$& $0.0\pm 0.2$\\
      $p(S>S^{\rm mod})$ & $ 0.45\pm 0.24 $ & $0.57\pm 0.20$ &$0.39 \pm 0.23 $ \\
      equiv. $\sigma$ & $0.8 \pm 0.4$ & $0.5\pm0.3 $&$0.9 \pm 0.4$ \\\hline
      \multicolumn{4}{c}{}\\

    \end{tabular}
  \end{center}
\end{table}

\begin{figure*}
  \centering
  \includegraphics[width=.55\linewidth]{{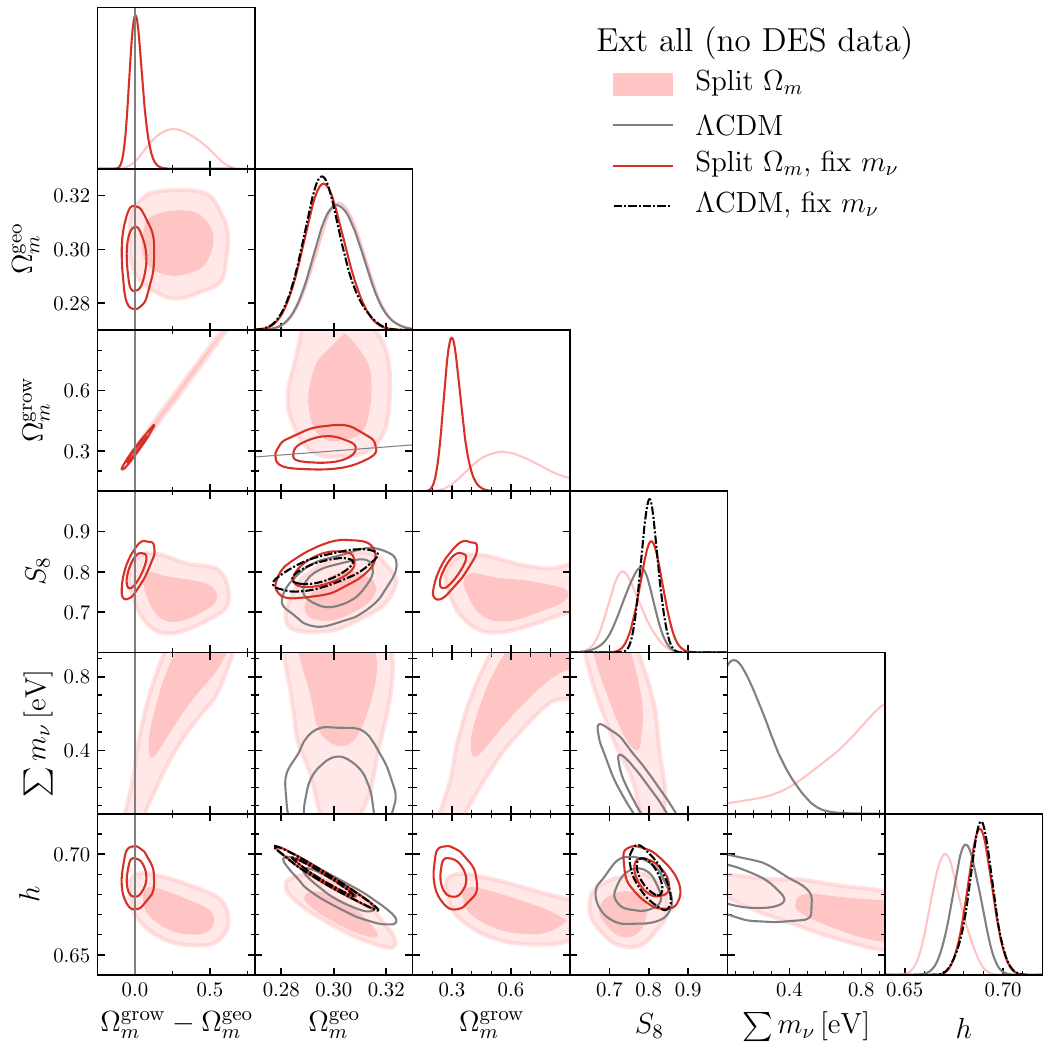}}
\caption{Ext-all constraints (not including any DES information) showing the effect of fixing neutrino mass for both the split \om parameterization and \lcdm.}
\label{fig:numodcomp_extall}
\end{figure*}

Additionally, \fig{fig:numodcomp_extall} shows how the Ext-all constraints change when we either fix \mnu to its minimum allowed value or we revert to \lcdm with $\omgrow=\omgeo$. In that Figure the shaded pink contours, which are the same as those in \fig{fig:degeneracies_sOm_p5br},  show our baseline DES+Ext-all constraints when \om is split and the sum of neutrino masses is varied. The solid red contours show how these constraints change when we fix $\mnu=0.06\,\text{eV}$, while the solid gray contours show what happens when we switch to \lcdm (but still vary neutrino mass). The dashed black lines show what happens when we both require $\omgrow=\omgeo$ and fix \mnu.

\bibliography{ggsplit_y1}{}

\label{lastpage}
\end{document}